\documentstyle[12pt,aaspp4]{article}
 
\slugcomment{}
 
\lefthead{Miller, et al.}
\righthead{Abell Cluster Correlations}
 
\begin{document}
 
\title{Projection, Spatial Correlations, and Anisotropies in a Large and Complete
Sample of Abell Clusters}

\author{Christopher J. Miller, David J. Batuski, Kurt A. Slinglend}
\affil{Department of Physics and Astronomy, University of Maine,
    Orono ME 04469-5709}
 
\and
 
\author{John. M. Hill}
\affil{Steward Observatory, University of Arizona, Tucson AZ 85721-0065}
 
\begin{abstract}
An analysis of $R \ge 1$ Abell clusters is presented for samples containing
recent redshifts from the MX Northern Abell Cluster Survey. The newly obtained
redshifts from the MX Survey as well as those from the ESO Nearby Abell Cluster
Survey (ENACS) provide the necessary data for the largest magnitude-limited
correlation analysis of rich clusters in the entire sky (excluding the galactic plane)
to date.
The MX Survey,
undertaken primarily for large-scale structure studies,
has
provided a large compilation of galaxy redshifts in Abell cluster fields
for the examination of
projection effects within the Abell catalog. 
In addition, a large and complete set of cluster velocities
(to $m_{10} = 16.8$) 
is used to determine
the two-point spatial correlation function and examine previously
suggested line-of-sight cluster selection bias within the Abell catalog.

The two-point spatial correlation function is
determined for a complete magnitude-limited (m$_{10} \le 16.8$) sample of
northern Abell clusters ($N_{cl} = 198$), a complete whole-sky (excluding the
galactic plane) magnitude-limited  (m$_{10} \le 16.8$) sample of Abell/ACO clusters
($N_{cl} = 289$), and a complete
whole-sky magnitude and volume-limited (m$_{10} \le 16.8$, $z \le 0.1$) sample of Abell/ACO
clusters ($N_{cl} = 238$)
In addition, we examined the largest sample of
cD clusters ($N_{cl} = 104$) in the Northern hemisphere.
We find $19.4 \le r_0 \le 23.3 h^{-1}$Mpc,
$-1.92 \le \gamma \le -1.83$
for the different cluster subsets examined (including the cD clusters). 

We have used the largest rich cluster data set available to
date to look for line-of-sight anisotropies within the Abell/ACO catalogs.
An examination of the correlation function
separated into its line-of-sight and perpendicular-to-line-of-sight
components show that the strong
anisotropy present in  previously studied Abell cluster datasets
is not present in our samples.
There are, however, indications of residual anisotropies
which we show are the result of two elongated superclusters, Ursa Majoris
and Corona Borealis whose axes lie near the line-of-sight.
After rotating these superclusters so that their semi-major
axes are perpendicular to the line-of-sight, we find no indication of
anisotropy in $\xi(\sigma,\pi)$.
The amplitude and slope of the correlation function remain the same
before and after these rotations. We also remove a subset of $R = 1$ Abell/ACO
clusters that show sizeable foreground/background galaxy contamination and again
find no change in the amplitude or slope of the correlation function.
We conclude that the correlation length of $R \ge 1$ Abell clusters
is not artificially enhanced by line-of-sight anisotropies.

\end{abstract}

\newpage
\section{Introduction}
The study of large-scale structure in the Universe plays a vital role in
the determination of the cosmological parameters and scenarios that 
describe the Universe from just after its creation to what we see today. It is the
visible structure on scales greater than $\sim 50 h^{-1}$Mpc (where $h$ is the Hubble constant
in units of 100 km s$^{-1}$ Mpc$^{-1}$) that can tell us directly about the
nature of the initial conditions  which could have generated such structures.
Most 
researchers believe that  
low amplitude Gaussian
density perturbations, 
resulting from quantum fluctuations in the earliest stages
of cosmic evolution, seeded the large-scale structure we see today 
(Guth \& Pi 1982; Linde 1982; Bardeen, Steinhardt \& Turner 1983).
(Although non-Gaussian initial conditions have been 
suggested, including topological defects (Zeldovich 1980; Vilenkin 1981)
and global structures (Turok 1989).)

While the data still remain sparse, 
there have been a variety of tracers used successfully 
for observing large-scale structure through redshift studies:
individual galaxies in wide-area magnitude-limited and pencil-beam surveys,
visually-selected galaxy clusters, plate-scanned clusters,
X-ray-selected clusters, groups of galaxies, and cD clusters.
On the largest of scales, galaxy clusters are more efficient 
tracers than individual galaxies.
While galaxy surveys such as the Sloan
Digital Sky Survey (SDSS) and the Australian 2dF Survey will undoubtedly  yield
excellent datasets for structure analyses, current surveys such as
the CfA (Geller \& Huchra, 1989) and
the Las Campanas Survey (Shectman et al. 1996)
among others, lack the depth, volume and completeness of coverage
needed for accurate evaluations of the two-point correlation function, power
spectrum, and other measures of structure on scales of $100 h^{-1}$Mpc and
greater. 
As of the writing of this paper,
both the SDSS and the 2dF surveys are years from completion and public access
to the data.

X-ray-selected cluster catalogs avoid visual selection or projection 
biases. The thermal X-ray emission from intracluster gas confined
to the gravitational potential wells of galaxy clusters is typically bright
and highly peaked at the gravitational center of the clusters. 
The ROSAT Brightest Cluster Sample (BCS) is a 90\% complete, flux limited sample
of 199 X-ray brightest clusters in the Northern Hemisphere that has provided
a detailed evaluation of the X-ray luminosity function (XLF) out to
a redshift $z \le 0.3$ (Ebeling, {\it et al.} 1997). This new
survey may provide an excellent
sample for large-scale structure studies and for comparison with optically
identified clusters.
However, there are questions as to exactly what type of
clusters (specifically in richness and shape) the X-ray samples trace. Bahcall \& Cen (1994)
show that flux-limited X-ray samples contain poor nearby clusters and rich distant clusters. 
A flux-limited sample of clusters
would contain a mixture of rich and poor clusters, 
both of which could be
incomplete in regions of the observed volume. 
Struble \& Ftaclas (1994) indicate that rich clusters are intrinsically more
nearly spherical than poor $R=0$ clusters. 
A preferable sample would contain the same type of cluster (i.e. clusters with similar
properties in shape and membership),
showing a constant average density in a large volume of space.

The first northern sky catalog of rich galaxy clusters was created by George Abell in
1958 from visually scanned sky survey plates. 
Abell's (1958) $R \ge 1$ clusters are ideally rare for large-scale structure
studies. They have an average spatial separation of  $\sim 50 h^{-1} $ Mpc,
making them very efficient samplers of the mass distribution on scales larger
than 100  $h^{-1} $ Mpc. A minimum of telescope time is required to map such
large scales with clusters, since spectroscopy of only a few relatively bright
galaxies is needed to obtain a reliable redshift for a cluster.
For the above
reasons, we have chosen the $R \ge 1$ Abell clusters as our sparse
tracers,
assuming that they represent the underlying mass distribution.

In the Northern
Hemisphere,
several researchers have used optically-observed rich
galaxy clusters to trace large-scale structure (for example, 
Bahcall \& Soneira 1983,
Batuski \& Burns 1985 and Postman, Huchra, \& Geller 1992). These efforts
provide evidence for a
strong signal of structure on scales $> 50  h^{-1} $ Mpc. However, the samples
utilized by these authors suffer from two limitations: small sample size
and single redshifts for many clusters. These limitations 
impose lower confidence in any of the statistical
analyses performed. 
Prior to the MX Northern Abell Cluster Survey,
the complete, magnitude-limited sample of Abell cluster redshifts
extended only to m$_{10} \le$ 16.5, where m$_{10}$ is the apparent
magnitude (photored) of the tenth brightest galaxy within the cluster.
This sample, presented by
Postman, Huchra and Geller (1992, hereafter PHG) consisted of
156 $R \ge1$ Abell clusters with $\delta \ge -27.5^{\circ}$.
When a galactic latitude limit of $|b| \ge 30^{\circ}$
is applied, 126 clusters remain in the sample.
Of these 126, 35 cluster velocities were determined by measuring the redshift of a single
galaxy within the respective cluster, resulting in unreasonably
large uncertainty in the velocities due to projection.
The small fraction of $R \ge 1$ clusters with observed redshifts, and even smaller
fraction with multiple-galaxy-determined redshifts, has
resulted in a severely limited sample for large-scale structure studies.

The MX Northern Abell Cluster Survey was undertaken in an attempt to rectify
these two incompleteness problems associated with the Abell sample of
clusters: the small number of clusters with multiple-galaxy observed redshifts
(that comprise a complete sample) and projection
effects on cluster membership.
The MX Survey has increased the latitude-limited, magnitude-limited sample size
to 198  $R \ge 1$ clusters (98\% complete to m$_{10} \le 16.8$).
At the same time, we have
reduced cluster redshift uncertainties, since 187 of these clusters now have more
than one measured cluster member redshift.
Slinglend {\it et al.} (1998) have obtained an average of nine cluster member redshifts in 88
observed clusters. The MX Survey has provided a substantially enlarged dataset of
$R \ge 1$ Abell clusters for the analyses in this paper. 
We remind the reader that the $R = 0$ subset of Abell clusters does not comprise a
statistically complete sample.  The $R =0$ clusters were included in the Abell catalog
only to enhance its value as a cluster finding list (Abell 1958; Struble \& Rood 1991), not
for use in spatial analyses.

The Abell and Abell, Corwin, and Olowin (1989-hereafter ACO) catalogs are the only optical cluster catalogs available that cover the entire
sky. The volume of space that is spatially analyzed in this paper is the largest to date.
If structures exist on scales greater than 50$h^{-1}$Mpc, any statistical analysis must use
volumes of space
substantially larger than the size of 
the largest features. Smaller surveys, such as all current galaxy surveys,
pencil-beam surveys and the plate-scanned cluster surveys ({\it e.g.} APM) are only
marginally large enough to
detect very large-scale structure. While the potential of the Abell/ACO
catalogs is obvious, their utility
has come under question due to anisotropies detected within the catalogs 
(Sutherland 1988; Efstathiou {\it et al.} 1992; Dalton {\it et al.} 1994; Peacock \& West 1992).
The analyses used in describing these anisotropies
indicate that Abell (1958) preferentially cataloged clusters near the 
lines-of-sight towards other clusters. Unfortunately, the samples used for
these analyses are themselves small and incomplete.
Incomplete datasets
used in large-scale structure analyses are subject to observational bias.
The Sutherland  and Peacock \& West  anisotropy analyses used 
highly incomplete samples. The PHG statistical sample of 208 $R \ge 0$ Abell clusters
(used by  Efstathiou {\it et al.} to show anisotropy within the catalog) is
only compete to m$_{10} = 16.5$ and contains over 100 $R =0$ clusters.
While there have been a number of papers suggesting
that the anisotropy is actually the result of high peculiar velocities, the elongation of
cluster pairs in the redshift direction, real geometric elongation of superclusters, or
some combination of these, none has offered definitive observational evidence (Jing {\it et al.} 1992; 
Bahcall {\it et al.} 1986).

The aim of this paper is to show that the Abell/ACO $R \ge 1$ clusters form 
a large and complete dataset with minimal projection effects and
line-of-sight anisotropies. The amplitude and
slope of the two-point spatial correlation function are re-examined using the largest (in number and
in volume) $R \ge 1$ clusters sample to date. 
In Section 2 projection effects are examined, 
Section 3 defines some independent and partially independent samples 
within the entire dataset for analysis, and Section 4 covers the determination of the 
mean space density of rich clusters. In Section 5
we present the cluster-cluster spatial correlation function for the subsamples of 
clusters.
Section 6 discusses the analysis of anisotropies in the Abell catalog,
and our results are summarized in Section 7.

\section{Projection effects}

In past large-scale structure analyses (using clusters), many of the cluster distances were
determined by the observation of a single galaxy member (Bahcall \& Soneira 1983; PHG).
To examine the uncertainty inherent in using clusters with only one measured redshift,
we calculated velocity differences for cluster redshifts published in PHG and Quintana
\& Ramirez (1995) when compared to the MX Survey redshifts of Slinglend {\it et al.} (1998).
We used 36 clusters within
$z \le 0.13$, all of which had only one measured galaxy before the MX Survey and which
now have an average of nine galaxy redshifts of cluster members. Figure 1 shows these velocity differences
as a function of redshift. A0121 and A0769 show the most extreme differences
of $\sim$ 14000 km s$^{-1}$ and $\sim$ 6000 km s$^{-1}$ respectively.
Five of the thirty-six clusters studied have velocity differences of $\sim 2000$ km s$^{-1}$
or greater,
while thirteen differ by $\sim$ 500 km s$^{-1}$.
We
conclude that projected membership 
(or at least extreme outliers in the velocity distribution of clusters) was 
a problem for 14\% of the previously single-measurement
clusters. The velocity errors
of $\sim 500$ km s$^{-1}$ for the second category of deviations (in 36\% of the clusters)
are the result of typical velocity dispersions within Abell clusters
which could
impact the structure signal on smaller scales ($\sim 5 h^{-1}$Mpc). 
However, the peculiar velocities of clusters have been shown to be $\sim 500$ km s$^{-1}$ 
(Bahcall \& Oh 1996),
so the use of the Hubble flow to study relative cluster positions on such small scales
is inappropriate in any case. 
Nevertheless, the results for the clusters in the higher deviation category
indicate that projection 
(or selection of a galaxy in the wings of the cluster velocity distribution)
can be problematic when using a sample that has a substantial number
of cluster redshifts based 
on a single measurement.

Another concern of clusters in the Abell catalog is the misclassification
of richness levels.
There have been claims that projection effects on cluster richness
designations is a serious problem in the Abell catalog ({\it e.g.} Lucey {\it et al.} 1983).
Recently however,
two of the most extensive rich cluster surveys, the MX Survey (Slinglend {\it et al.}
1998) and the ENACS survey (Katgert {\it et al.} 1996), find only $\sim 10\%$
of clusters to have more than $\sim 30\%$ of galaxies per cluster field to
be foreground/background. 
Abell (1958) in fact,  claimed that up to 30\% of the galaxies
within a cluster field were background/foreground galaxies.
After subtracting  a locally estimated background count, 
Abell determined 
the total number of galaxies within each cluster to his specified magnitude limit.
Clusters containing 30-49 galaxies 
were classified as $R = 0$, clusters containing 50-79 galaxies were classified as $R = 1$,
and so on. The difficulty for Abell was twofold: (1)
not knowing the redshifts of any of the galaxies in fields
he was examining and (2) having to estimate the redshifts of clusters as well
as their linear diameters on the sky (corresponding to $1.5h^{-1}$Mpc).
Difficulty (1) was addressed by subtracting the background galaxy count
(derived from a nearby region on the plate apparently `free' of clusters). Difficulty (2)
was handled by using an m$_{10} -z$ relation derived from 18 clusters with known
redshifts (in 1958). 
The 22 galaxies (on average) per cluster field observed by the MX Survey provides an excellent
database for comparing the true background fraction with that estimated by Abell.
In addition, we can examine how many clusters had underestimated (or overestimated) angular diameters
which would affect their richness assignment.
We note that the MX Survey has not provided enough galaxy redshifts for a 
determination of the absolute cluster membership of Abell clusters. 

We examined galaxies within one  
Abell radius of the cluster center as defined using an m$_{10} -z$ derived from
Abell (1958). The estimator is
\begin{equation}
log(cz) = 0.71 + 0.23m_{10}.
\end{equation}

We first look at those clusters for which the actual redshift and the estimated
redshift differ by less than 4000 km s$^{-1}$, {\it i.e.} where Abell's distance
estimate was `correct'. There are 29/80 such cases and we find that
27\% of the galaxies observed in these fields were foreground/background.
Therefore, when Abell's redshift estimate was near the true redshift of the cluster,
subtracting up to $\sim 30\%$ of the galaxy count due to foreground/background contamination
is appropriate.
For the remaining
clusters, we find that the majority of the cluster redshifts
(44/80) are over-estimated using equation (1), 
resulting in a much smaller effective radius within which galaxies were counted. Since most Abell 
clusters have well-defined cores, we expect the fraction of foreground/background galaxies
to be lower for these cases. In fact, we find only 20\% contamination when the estimated 
cluster redshift is much greater ($> +4000$ km s$^{-1}$)
than the true redshift.
According to Abell (1958), altering the angular diameter corresponding
to one Abell radius ($1.5 h^{-1}$Mpc) by $\pm{30\%}$  would not affect the richness classification.
However, we find that 37 of the 44 clusters with overestimated redshifts
have angular diameters (corresponding to one Abell radius)
on the sky which are more than 30\% larger than that used by Abell.
Therefore, 91\% of the clusters
in this analysis (m$_{10} \le 16.8, R \ge 1$)
would have the same or greater richness classification had
their true redshifts been used.
Only 7/80 clusters have underestimated redshifts according to 
equation (1) . 
These results do not change significantly ($1 - 3\%$)
when using Abell radii up to 2.5  h$^{-1}$Mpc.

The MX Survey has provided too few galaxy redshifts per cluster field
for an accurate galaxy-member count.
However, by examining the true redshifts of the clusters and their 
foreground/background galaxy fractions,
we can conclude that the apparent richnesses, as classified
by Abell (1958), are generally accurate (or perhaps underestimated). Since the remaining analyses
in this paper are based on $R \ge 1$ clusters, underestimated richness levels would not
alter any of the results. 

\section{The Cluster Samples}
The addition of the MX Survey clusters to ones previously measured
by other researchers (PHG, Struble \& Rood 1987)
brings the total number of observed clusters with
m$_{10} \le 16.8$, $R \ge 1$, $ -17^{\circ} \le \delta \le 90^{\circ}$ to 193 (out
of 198 listed in the Abell (1958) catalog).
A recent survey in the southern hemisphere, the ESO Nearby Abell Cluster Survey
(ENACS)
has added redshifts for 104 $R \ge 1$ ACO
clusters with mean
redshifts $z \le  0.1$ (Katgert {\it et al.} 1996). The depth of the ENACS survey is
similar to the 
MX survey and thus provides an excellent southern dataset that can be combined
with the MX survey for all-sky studies.
We note that there are differences between $R \ge 1$ Abell clusters
and $R \ge 1 $ ACO clusters ({\it e.g.} Batuski {\it et al.} 1989 and references therein).
Such differences between the catalogs include: 
$V$ {\it vs.} $R$ band magnitude determinations (ACO and Abell respectively),
greater sensitivity of the IIIa-J
plates used in the ACO catalog, and a
global determination of the background galaxy count vs. local determinations
(ACO and Abell respectively).
For the correlation analyses presented in this paper we will be
examining four different subsets of clusters summarized below.

\begin{enumerate}
\item{\bf The northern magnitude-limited sample}
is the 198 Abell m$_{10} \le 16.8$, R$ \ge 1$ clusters with
$0^{h} \le \alpha \le 24^{h}$, 
$-17^{\circ} \le \delta \le 90^{\circ}$, and $|b| \ge 30^{\circ}$. 
Five of these clusters have estimated redshifts from the Batuski \& Burns (1985)
magnitude-redshift relation.

\item{\bf The whole-sky magnitude-limited sample} contains the 198 Abell m$_{10} \le 16.8$, R$ \ge 1$
clusters in the above northern sample plus 91 $R \ge 1$,  m$_{10} \le 16.8$,
ACO clusters with $-90^{\circ} \le \delta \le -17^{\circ}$, $0^{h} \le \alpha \le 24^{h}$.
Six of the ACO clusters have estimated redshifts according to the m$_{10} - z$ relations
in the ACO the catalog.
In section 4, we find a higher mean density for the ACO sample of clusters compared to
the Abell sample, supporting previous claims that the ACO clusters are slightly poorer than their
northern counterparts. Therefore, to create a homogeneous sample containing both ACO and Abell clusters
(to similar richness),
we have excluded the 14 poorest ACO $R =1$ clusters (those with N$_{gal} \le 54$ where N$_{gal}$ is
the number of galaxies within an Abell radius as listed in the ACO catalog). The remaining sample of
91 ACO clusters has a mean spatial number
density similar to that of the Abell clusters (see section 4). 
This sample is the largest $R \ge 1$ cluster sample to date and contains 289 clusters.

\item{\bf The whole-sky statistical sample} is a subset of the previous sample
excluding
any clusters that lie beyond $z = 0.10$.
The northern Abell cluster magnitude and volume-limited sample
contains 156 clusters (two with estimated redshifts).
The Southern hemisphere magnitude and volume-limited sample contains
82 ACO clusters (five with estimated redshifts). 

\item{\bf The volume-limited cD sample}
contains cD clusters as defined in ACO (1989). This
sample is volume-limited to  $D \le 350$ h$^{-1}$Mpc, and 
$0^{h} \le \alpha \le 24^{h}$, $-17^{\circ} \le \delta  \le 90^{\circ}$,
and $|b| \le 30^{\circ}$. 
We apply no richness constraint to this sample of 104 clusters.
A smaller sample of cD clusters ($N_{cl} = 64$) was previously used for large-scale structure analyses in
West \& van den Bergh (1991). Schombert \& West (1990) have shown that
cD clusters have a distribution representative of Abell clusters in
general. Thus by limiting a sample to only cD clusters, we should not
be introducing any additional biases.
This sample is somewhat independent from the others, since it 
has different selection criteria and
contains a substantial number of $R = 0 $ clusters. 
\end{enumerate}

Distances to all clusters were determined from
Sandage (1975) where
\begin{equation}
D = \frac{cz}{H_o}{\frac{(1+\frac{z}{2})}{(1+z)}}
\end{equation}
for a Friedman universe with $q_0 = 0$.
All of the samples have an additional constraint of $|b| \ge 30^{\circ}$, to
limit the effects of 
galactic obscuration near the plane of the Galaxy.
Figure 2 shows Aitoff projections for the northern magnitude-limited
sample and the whole-sky magnitude-limited sample (in galactic coordinates).

The number of clusters within the whole-sky sample makes it 
the largest set of
$R \ge 1$ Abell clusters that has yet been used for correlation analyses.
In addition, the volume traced by these clusters ($\sim 5\times 10^{7} h^{-3}$Mpc$^{3}$)
is the largest using a complete sample. Together, the large number of clusters and volume of
space studied in the following analyses provide a most complete picture of
large-scale structure in the local Universe.

\section{The Spatial Number Density of Clusters}
The number density in a proper volume, $n_p(z)$, of a survey
can be calculated for a Friedmann Universe (Narlikar 1983) :
\begin{equation}
dn_o = d\Omega({\frac{c}{H_o}})^3
{\frac{\{q_0z+(q_0-1)[(1+2zq_0)^{1/2}-1]\}^2n_pdz}
{q_0^4(1+z)^6(1+2q_0z)^{1/2}}},
\end{equation}
and we have 
\begin{equation}
dn_o = d\Omega({\frac{c}{H_o}})^3
{\frac{z^2(2+z)^2n_pdz}{4(1+z)^6}}
\end{equation}
for $q_0 = 0$, where $dn_o$ is the number of observed clusters
within dz and $d\Omega$ is the solid
angle subtended by the survey.
 
If we include an unknown selection function in z, $S(z)$,
and the evolution of $n_p$ via $E(z)$, we must let
$n_p(z) \rightarrow n_p(z)S(z)E(z)$ (Kolb \& Turner 1990).
For the samples mentioned above, we assume $E(z) = S(z) = 1$ for
simplicity, since evolution of the proper number density of clusters
should be negligible at these low redshifts and we are only studying a volume that 
has a nearly constant density (see Figure 3), making S(z) $\sim 1$. 
Figure 3 shows the number density ($n_p$) as a function of
z for the northern magnitude-limited sample and the southern magnitude-limited
sample. 

The solid angle of the sky covered by the MX Northern sample of 198 clusters
is 4.25 steradians. The
whole-sky magnitude-limited sample of 282 clusters covers an area of 
6.28 steradians (the
entire sky excluding galactic latitudes below $30^{\circ}$.)
Using Figure 3, we see the number density no longer being
constant at $z = 0.09$. 
Averaging the weighted mean number densities in Figures 3a,b (to $z = 0.09$)
we find
$\bar{n}_p = 6.6 (\pm 0.6) \times 10^{-6} h^{3}$Mpc$^{-3}$
for the northern Abell clusters and $ \bar{n}_p = 7.6 (\pm 0.9) \times 10^{-6} h^{3}$Mpc$^{-3}$
for the southern ACO clusters (using $1 \sigma$ errors).
To create the whole-sky clusters sample, we excluded 14 ACO clusters (those with N$_{gal} \le 54$),
reducing the density for the ACO sample to $ \bar{n}_p = 6.5 (\pm 0.9) \times 10^{-6} h^{3}$Mpc$^{-3}$.
[The MX Survey II 
is nearing completion after observing most Abell clusters with $16.9 \le m_{10} \le 17.0$
(Miller {\it et al.} 1999). The additional $\sim$ 100 clusters provided by the MX Survey II should
increase the density slightly more, as a few low redshift clusters are added. 
With the results from the MX Survey II, we expect to 
have a sample deep enough in magnitude to calculate an accurate density for a 
volume-limited sample of clusters to $ z = 0.10$.]

The number density for the northern clusters
implies a mean separation length of $\sim 50 h^{-1}$Mpc.
For comparison, the
spatial number density of APM clusters is three times as large at 
$\sim 2.4 \times 10^{-5}  h^{3}$Mpc$^{-3}$. APM 
clusters are thus even less rich than $R=0$ Abell clusters and therefore much less
rare (Bahcall \& West 1992).
Our redshift limit of $z = 0.10$ for the magnitude and
volume-limited samples
(discussed in Section 3)
was based on the redshift where $n_p$ differs from $\bar{n}_p$ by more than
2$\sigma$ in Figure 3. This occurs for both the northern
and southern clusters at $z \sim 0.10$. 

\section{The Cluster Spatial Correlation Function}
We use the following estimator derived in Hamilton (1993) for
the determination of the correlation function: 
\begin{equation}
\xi(r) = \frac{DD(r) \times RR(r)}{DR(r)^2} -1,
\end{equation}
where $DD$, $RR$, and $DR$ are the data-data, random-random and 
data-random paircounts respectively
with separations between $ r - \frac{\Delta r}{2}$ and $ r + \frac{\Delta r}{2}$.
We refer the reader to Hamilton (1993) and Landy \& Szalay (1993)
for an analytical analysis of the 
estimator.
Compared to previous estimators (see Bahcall \& Soneira (1983) and PHG), this one is proposed
to be less affected by
uncertainties in the mean number density where separations are
large and $\xi$ is small.
Recently, Ratcliffe {\it et al.} (1998) used N-body simulations to show that
the Equation (5) provided the most accurate results when compared
to other estimators. 

The random paircounts (DR, RR) are evaluated by averaging over 400
catalogs generated with the same number of pseudo-clusters as the
sample under consideration. The angular coordinates in these catalogs
are randomly assigned
with the same boundary conditions as the survey and with a galactic
latitude selection function,
\begin{equation}
P(b) = 10^{\alpha(1-csc|b|)},
\end{equation}
with $\alpha = 0.32$ imposed to account for residual galactic obscuration
after the latitude cut at $|b| = 30^{\circ}$ (PHG, Peacock \& West 1992).
The redshifts assigned to the random catalog points are 
selected from the observed data after being smoothed
with a Gaussian of width 3000 km s$^{-1}$. This technique
corrects for radial density gradients on small scales in the
observed distribution.
For the whole-sky sample, the random points with $\delta \ge -17^{\circ}$ are assigned
redshifts from the Abell clusters, while the points with  $\delta \le -17^{\circ}$ are assigned
redshifts from the ACO catalog.
We have chosen to limit the cluster redshifts in the Whole-sky sample
to $z = 0.10$ where the average densities begin to drop off dramatically. However,
by using the 
observed redshifts for our random
catalogs, we are effectively modeling any selection effects in $z$, making the chosen
limit of  $z = 0.10$ somewhat arbitrary. 

The spatial correlation function for luminous mass concentrations
has been shown by various authors to have the following 
power-law form:
\begin{equation}
\xi(r) = (\frac{r}{r_0})^{\gamma}.
\end{equation}
For galaxy-galaxy spatial correlations, $r_0 \approx 5 h^{-1}$Mpc and
$\gamma = -1.8$ (see Willmer {\it et al.} (1998)
for a review of galaxy-galaxy correlation function results).
However, cluster-cluster correlation results 
have been more uncertain (see Table 1.)
Bahcall \& Soneira (1983) originally found $r_0 \approx 24 h^{-1}$Mpc
and $\gamma = -1.8$. However, 
the small size of this sample, 104 $R \ge 1$ clusters, 
and its use of many single-redshift clusters leave considerable 
uncertainty in those results.
PHG later found similar results for $\gamma$ and
$r_0$  with a slightly larger
sample of 156 $R \ge 1$ clusters. We reexamined this sample, with 
a galactic latitude cut of $|b| \ge 30^{\circ}$ to minimize any
effects in the Abell catalog due to obscuration by the Galaxy. 
The final sample of 126 clusters
is very similar to the Bahcall \& Soneira (1983)
sample and we expect the same results (and large uncertainties). 
Our tests produced results nearly identical to those reported by PHG 
(Table 1), 
although we would not have been surprised to see 
small differences as a result of a few different redshifts for
clusters that had previously had only one measured redshift (before the MX survey).

Soon after the
PHG results, Peacock and West (1992) compiled a volume-limited (to $z = 0.08$)
sample of 195 $R \ge 1$ Abell and ACO clusters and found a reduced $r_0 = 21  h^{-1}$Mpc and
increased $\gamma = -2.0$.
It should be noted that
Peacock \& West did not use a magnitude limit for their sample and the data was not
complete beyond m$_{10} \ge 16.5$, although many clusters with m$_{10} > 16.5$
were used. It is highly probable that the m$_{10} \ge 16.5$ subset of clusters used in the
Peacock \& West sample was not a fair sampling. Many dimmer clusters were observed
for a specific reason (such as being a supercluster
member, particularly rich, containing substructure, etc).
The use of such incomplete datasets raises the risk of detecting spurious
structure due to sampling effects.
There have been other determinations of 
$r_0$ and $\gamma$ for Abell clusters via cD clusters and X-ray clusters
selected from within the Abell catalog with results $16 \le r_0  \le 22 (h^{-1}$Mpc)
and $-1.9 \ge \gamma \ge -1.7$ (West \& van den Bergh 1991;
Nichol {\it et al.} 1994).

With the recent evidence for redshift-versus-sky-position inhomogeneities in
samples of Abell clusters (presumably from the visual selection of clusters) 
(Sutherland 1988; Efstathiou {\it et al.} 1992), 
several groups have 
conducted redshift surveys of catalogs of clusters produced 
through machine-scanning of sky-survey
plates, {\it e.g.} the APM galaxy survey (Maddox {\it et al.} 1990a,b) and
the EDSGC cluster catalog (Lumsden {\it et al.} 1992).
The correlation length for clusters selected from the APM catalog show
$13 \le r_0$ ($h^{-1}$Mpc) $ \le 15$ and $\gamma \approx -2$
(Efstathiou {\it et al.} 1992; Dalton {\it et al.} 1994). 
These determinations are based on samples with large numbers (up to 364) 
of more common, less rich clusters 
within an area on the sky of 1.3 steradians.
The correlation length for the EDSGC clusters was reported to
be $r_0 = 16.2 \pm{2.3}$ with $ \gamma = -2.0 \pm{0.2}$ (Nichol {\it et al.} 1992). 
A list of results for the slope and correlation length for the
various determinations of $\xi$ are listed in Table 1.

The large difference between the Abell cluster correlation length and that of the
plate scanned cluster catalogs has been the subject of much debate.
Unfortunately, until this work, the samples of Abell clusters analyzed 
were either small with many clusters having only one measured galaxy redshift (Bahcall
\& Soneira 1983),
incomplete in magnitude (Peacock \& West 1992), or contained $R =0$ clusters (PHG).
This work eliminates all of these problems with a large magnitude-complete sample
of $R \ge 1$ Abell/ACO clusters. We present our two-point correlation function results
in the next section.

\begin{deluxetable}{cccc}
\tablenum{1a}
\tablewidth{0pt}
\tablecaption{{\bf Results from Other Surveys}}
\tablehead{
\colhead{Reference} &
\colhead{Clusters} &
\colhead{$r_0 (h^{-1}Mpc)$} &
\colhead{$\gamma$}
}
\startdata
Bahcall \& Soneira 1983 & 104 & 25.0 & -1.8 \nl
Ling {\it et al.} 1986 & 104 & 21.9$_{-5.1}^{+7.1}$ & -1.7$\pm{0.17}$ \nl
Huchra {it et al.} 1990 & 145 & 20.8$_{-6.9}^{+6.7}$ & -1.8 \nl
West \& van der Bergh 1991 & 64 & 22.1$\pm{6.8}$ & -1.7$\pm{0.5}$ \nl
Postman {\it et al.} 1992 & 156 & 23.7$_{-9.0}^{+7.9}$ & -1.8$\pm{0.2}$ \nl
Peacock \& West 1992 & 195 & 21.1$\pm{1.3}$ & -2.0$\pm{0.2}$ \nl
Efstathiou {\it et al.} 1992 & 211 & 14.0 $\pm{1.3}$ & -1.9 $\pm{0.23}$ \nl
Nichol {\it et al.} 1992 & 79 & 16.2 $\pm{2.3}$ & -2.0 $\pm{0.2}$ \nl
Dalton  {\it et al.} 1994 & 364 & 14.3 $\pm{2.5}$ & -2.05  $\pm{.12}$ \nl
Nichol  {\it et al.} 1994 & 67 & 16.1  $\pm{3.4}$ & -1.9  $\pm{3.4}$ \nl
Abadi {\it et al.} 1998 & 248 & $21.1^{+1.6}_{-2.3}$ & -1.92 \nl
 
\enddata
\end{deluxetable}
\nopagebreak
\begin{deluxetable}{ll}
\tablenum{1b}
\tablewidth{0pt}
\tablecaption{{\bf Notes}}
\tablehead{
\colhead{Reference} &
\colhead{Notes}}
\startdata
Bahcall \& Soneira 1983 & Abell Statistical Sample, R$\ge1, D\le4$ \nl 
Ling {\it et al.} 1986 & same data as above \nl
Huchra {\it et al.} 1990 & Deep Abell Survey, R$\ge1$ \nl
West \& van der Bergh 1991 & cD Abell Clusters \nl
Postman {\it et al.} 1992 & R$\ge1$, m$_{10} \le 16.5$ \nl
Peacock \& West 1992 & Volume limited, R$\ge1$ \nl
Efstathiou {\it et al.} 1992 & {APM\tablenotemark{a}} $R_{DEMS} \ge 20$ \nl
Nichol {\it et al.} 1992 & {EDSGC\tablenotemark{a}} $R_{EDSGC} \ge 22$ \nl 
Dalton  {\it et al.} 1994 & {APM\tablenotemark{a}} $R_{APM} \ge 50 $ where $R_{APM} = 24.8 + R_{DEMS}$ \nl
Nichol  {\it et al.} 1994 & X-ray selected $R \ge 0$ Abell clusters \nl
Abadi {\it et al.} 1998 & X-ray selected $R \ge 0$ Abell clusters \nl
\tablenotetext{a}{We refer the reader to the individual APM and EDSGC papers for clarification on the
richness levels for those catalogs. We simply note here that the clusters in both the APM catalog and
the EDSGC catalog are poorer than the $R \ge 1$ Abell clusters.}
\enddata
\end{deluxetable}

\subsection{Results for the MX Survey Sample} 

Figure 4 shows the results for the spatial correlation function for the four 
different samples defined in Section 3. 
The three lines on the graphs indicate two
enveloping limits that span the results in Table 1
($r_0 = 25$, $\gamma = -1.8$  and 
$r_0 = 15$ and $ \gamma = -2.0$) and the best power-law fit for $\xi(r)$ from  our data
points (from $ r = 4 h^{-1}$Mpc to $ 35 h^{-1}$Mpc). 
The error bars plotted in Figure 4 are
determined from
\begin{equation}
\delta\xi = \frac{(1+\xi)}{\sqrt{DD}}.
\end{equation}
However, we note that Croft \& Efstathiou (1994) have shown that this underestimates
the true error by a factor of
$1.3 \rightarrow 1.7$.
We have excluded the
bins of smaller separations ($r \le 4.5 h^{-1}$Mpc)
from the fit for the following reasons: (1) the number of data-data pairs at small separations
becomes unreasonably small; (2) we get nearer to the actual diameter of a rich
Abell cluster where projection effects compromise differentiation
from a foreground and background cluster; and (3)  the $\sim $500 km s$^{-1}$
cluster peculiar velocities mentioned in Section 2 will mask structure.
Notice that the amplitudes in Figure 4a-d are similar to that of
previous, smaller samples used by Bahcall \& Soneria (1983) and PHG, yet
only the cD clusters show strong correlations beyond $r \sim 25h^{-1}$Mpc.
These results, listed in Table 2, provide strong support for previous claims
of a large two-point correlation function amplitude. Yet at the same time,
we find only minimal correlations beyond $r \sim 25h^{-1}$Mpc for the northern
magnitude-limited sample and both whole-sky samples. We point out that using
an older estimator ({\it e.g.} $\frac{DD}{RR} - 1$) shows slightly larger
correlations out to $\sim 50 h^{-1}$Mpc before going negative.

The slopes and correlation lengths are similar for all the datasets examined,
including the cD cluster sample (which is significantly different from the other samples
since it contains $R =0$ clusters).
All four samples examined have 
$19.4 \le r_0 \le 23.4 (h^{-1}$Mpc) and $-1.92 \le \gamma \le -1.83$.
On large scales, we find that $\xi(r)$
becomes negative at $\sim 50 h^{-1}$Mpc, 
although the error on $\xi$ in this region is comparable to $\xi$ itself.
The increased number of redshifts, in addition to the increased accuracy 
of those redshifts due to multiple-galaxy observations, has produced a much
improved fit (with much smaller errors) to the rich cluster correlation function.
These results, based on the largest
and most complete sample of rich clusters assembled to date, show 
that the correlation length for $R \ge 1$ Abell clusters is significantly
larger ($\sim 3\sigma$) than results for the APM survey.

\subsection{The Effects of Superclusters on $r_0$}

The effect of the Corona Borealis supercluster (Cor Bor) on the amplitude of
the correlation was illustrated in Postman {\it et al}
(1988) and again in PHG. Using the Bahcall \& Soneira (1983) $D \le 4$, $R \ge 1$
sample of clusters, Postman {\it et al.} (1988) concluded
that Cor Bor is responsible for $\sim 30\%$ of the amplitude of $\xi(r)$. They later reexamined
this effect using a larger sample of $R \ge 1$ clusters
and found that Cor Bor contributed slightly less, $\sim 20\%$ (Postman, Huchra and Geller, 1992).
Using the whole-sky
magnitude-limited sample of $R \ge 1$ clusters, we find that $r_0$
decreases by $\sim 8\%$
after excluding Cor Bor (A2061, A2065, A2067, A2079, A2089, A2092) (see Table 2).
We find a slightly larger reduction in $r_0$ after excluding the newly discovered dense and compact
supercluster in Microscopium. This supercluster consists of A3677, A3682, A3691, A3693, A3695 and A3705
and was first reported by Zucca {\it et al.} (1993) and 
many redshifts were observed by ENACS (Kagert {\it et al.} 1996).
The exclusion of the Microscopium Supercluster (MSC)
reduces $r_0$ by
$\sim 13\%$. Together, the removal of Cor Bor and Microscopium reduces $r_0$ by 20\% (from 
$23.4h^{-1}$Mpc to $18.4h^{-1}$Mpc) and steepens the
slope from $\gamma = -1.83$ to $\gamma = -2.13$.
It is important to note that this lower value of $r_0$ is within $1\sigma$ of the
value obtained for the APM cluster results (Efstathiou {\it et al.} 1992) and
that both Cor Bor and the MSC lie outside of the
right ascension and declination limits of the APM Survey.

While the effects
on the correlation function from individual superclusters seem to diminish
as we increase the sample size,
we are finding more of these dense and compact superclusters as we extend the survey
to greater depths.
In fact, Batuski {\it et al.} (1999) have recently discovered a collection of six
Abell/ACO clusters at $z = 0.11$ with a spatial number overdensity similar to that of the
Corona Borealis supercluster. Thus, including the Shapley Concentration, we now have four such dense, compact
superclusters in the local Universe.
The existence of these superclusters has a clear impact on the correlation
function amplitude and slope.
(We note that most of the Shapley Concentration does not appear in these analyses due to the
imposed galactic latitude limit of $|b| \ge 30^{\circ}$.)

\begin{deluxetable}{llll}
\tablewidth{0pt}
\tablenum{2}
\tablecolumns{5}
\tablecaption{{\bf Results for the Power-Law Fits of $\xi$}} 
\tablehead{
\colhead{Sample} &
\colhead{Size } &
\colhead{$\gamma$}  &
\colhead{$r_0$ } \nl 
\colhead{m$_{10}\le 16.8, R \ge 1$, $|b| > 30^{\circ}$} &\colhead{}
& \colhead{} & \colhead{($h^{-1}$Mpc)} \nl
}
\startdata
Northern, $\delta \ge -17^{\circ}$  & 198 & $-1.92\pm{0.22}$ & $19.42^{-5.32}_{+4.06}$   \nl
Wholesky  & 289 & $-1.83\pm{0.15}$ & $23.36^{-4.19}_{+3.66}$   \nl
Wholesky,  $z < 0.10$   & 238 & $-1.88\pm{0.16}$ & $22.79^{-4.58}_{+3.89}$  \nl 
Wholesky  & 283 & $-1.85\pm{0.19}$ & $21.36^{-4.13}_{+3.48}$ \nl 
~~~~ (excluding Cor Bor) & & & \nl
Wholesky   & 283 & $-1.94\pm{0.19}$ & $20.42^{-4.46}_{+3.69}$ \nl
~~~~ (excluding MSC)  & & & \nl
Wholesky   & 285 & $-1.82\pm{0.16}$ & $22.66^{-4.35}_{+3.77}$ \nl
~~~~ (excluding Ursa Majoris)  & & & \nl
Wholesky   & 277 & $-2.13\pm{0.22}$ & $18.38^{-4.13}_{+3.28}$ \nl
~~~~ (excluding Cor Bor and MSC)  & & & \nl
Wholesky   & 289 & $-1.90\pm{0.15}$ & $23.14^{-4.24}_{+3.62}$ \nl
~~~~ (after rotating Ursa Majoris and Cor Bor) & & & \nl
Wholesky   & 270 & $-1.94\pm{0.16}$ & $22.63^{-4.30}_{+3.65}$ \nl
~~~~ 19 \lq contaminated\rq clusters removed & & & \nl
Northern cD, Distance $ < 350 h^{-1}$Mpc $\delta \ge -17^{\circ}$ & 104 & $-1.88\pm{0.47}$ & 
$20.10^{-11.32}_{+9.86}$  \nl  
\enddata
\end{deluxetable}

\section{Anisotropy in the Abell catalog}
The amplitude of the two-point spatial correlation function for clusters has been
a controversial subject since the seminal work by Bahcall \& Soneira (1983). Soon
after their report that $r_0 \sim 25h^{-1}$Mpc for rich clusters, numerical
simulations based on the standard model ($\Omega = 1$, CDM) were conducted that 
failed to reproduce this high amplitude after normalization to the galaxy distribution
or the cosmic microwave background (Bardeen, Bond, Efstathiou 1987; White {\it et al.} 1987). 
In addition, analytical work by Kaiser (1984) showed that if clusters form at the
high density peaks of Gaussian random fields, then we expect the cluster correlation
function to approach the matter correlation function at large separations, and the
matter two-point spatial correlation function is predicted to become negative at
$\sim 20 ({\Omega}h^2)^{-1}$ in a CDM Universe (Davis {\it et al.} 1985). Yet
the observed two-point correlation function for rich clusters remains positive
out to $\sim 50h^{-1}$Mpc.

In response to the simulation and analytical results, Sutherland (1988) proposed that
the high amplitude of $r_0$ for the Abell catalog was the result of the spurious
enhancement of the number of cluster pairs at small separations.
`Corrective' techniques were created to account for these supposed
anisotropies which reduced the value of $r_0$. These techniques (see Sutherland (1988) or
Efsathiou {\it et al.} (1992)) typically subtracted out individual clusters or
cluster pairs that were too close together on the plane of the sky.
Other explanations for the anisotropy (excess pairs with small angular separations)
detected in the Abell catalog have
been suggested. Bahcall, Soneira \& Burgett (1986)
have proposed that the anisotropy could be 
caused by the finger-of-God effect or possibly supercluster elongation. 
Jing, Plionis, \& Valdarnini (1992) found that the elongation of cluster pairs
in the redshift direction occurred nearly as often (15\% to 30\% of the cases)
in model universes (with pseudo-clusters placed
near the peaks of a Gaussian density field) as in the Abell catalog.
Peacock \& West (1992) found that most of the anisotropies were in the
statistically incomplete subset of $R =0$ clusters.
We will show that the $R \ge 1$ subset of Abell clusters contains only a
residual amount of anisotropies which can be accounted for
by specific superclusters, elongated roughly along the
line-of-sight.

To search for this anisotropy, we examined $\xi(r)$ after separating $r$ into its line of sight component 
and the component perpendicular to the line of sight (Sutherland 1988):
\begin{equation}
r^2 = \sigma^2 + \pi^2
\end{equation}
where
\begin{equation}
\pi = |z_1 - z_2|
\end{equation}
Figure 5 shows contour plots of $\xi(\sigma, \pi)$ over the range
$0 - 100 h^{-1}$Mpc for both
$\sigma$ and $\pi$, in 10 $h^{-1}$Mpc bins. The heavy contour
line is $\xi(\sigma, \pi) = 1$, indicating relatively strong correlations.
These types of plots have been used by numerous authors as an indicator of
line-of-sight anisotropy within cluster catalogs (Peacock \& West 1992; Efstathiou
{\it et al.} 1992; Nichol {\it et al.} 1992; Dalton {\it et al.} 1994).
However, a review of the literature shows some confusion in the interpretation
of these plots. Therefore, we will first examine the PHG dataset and discuss why
(and where) it has  
been shown to be anisotropic.

Figure 5a recreates  $\xi(\sigma, \pi)$ for
the PHG statistical sample. This sample was used by
Efstathiou {\it et al.} (1992) to contrast with the APM sample.
The reader will notice a region of strong correlation with
$\sigma \le 20  h^{-1}$Mpc  and $50 \le \pi \le 80 h^{-1}$Mpc. The most
probable explanation for these correlations is the preferential selection of
clusters nearby on the plane of the sky with relatively large separations in redshift.
This indicates that some of the clusters, within the pairs that are causing
these high correlations, should have been classified with a lower richness
(and possibly not even  $R \ge 0$ clusters at all). We must also recognize
the level of incompleteness in the $R =0$ subset of Abell clusters as a contributor
to these anisotropies, although it is difficult to say exactly what effect
this would have. Strong correlations in
this region of $\sigma$ and $\pi$ 
indicate observational bias within the sample
examined. Peacock \& West (1992) noticed this anisotropy in a volume-limited
sample of $R =0$ Abell clusters and recommended that $R =0$ clusters
not be used in statistical analyses.
The reader will also notice the area of strong correlations with
$\sigma \le 20  h^{-1}$Mpc  and $ \pi \le 40  h^{-1}$Mpc. This
2:1 elongation of $\xi(\sigma, \pi)$ in the strong clustering region can be explained
by a number of effects: high cluster peculiar velocities ($\sim 1500$ km s$^{-1}$),
a large number of pairs with small angular separations and moderate redshift
separations, or real geometric elongation of superclusters with multiple
cluster members distributed along the line-of-sight. While line-of-sight
anisotropy is also an explanation, it is  not the only possibility.

Contour plots, such as Figure 5, have been widely used to compare 
machine plate-scanned surveys with the Abell catalog (Efstathiou {\it et al.} 1992;
Nichol {\it et al.} 1992; Dalton {\it et al.} 1994).
The conclusion from these analyses is that plate-scanned catalogs 
(such as the APM and EDSGC catalogs) do not suffer from the anisotropy
as seen in the PHG Abell cluster sample (Figure 5a).
Results from APM cluster analyses by Efstathiou {\it et al} and
Dalton {\it et al.}  show no correlations for
small $\sigma$ and large $\pi$ ($50 \le \pi \le 80 h^{-1}$Mpc), yet
there is a similar 2:1 elongation of $\xi(\sigma, \pi)$
in the strong clustering region. In contrast to the APM catalog,
the Nichol {\it et al.} analysis of the
EDSGC cluster catalog 
reports no elongation in the strong clustering region. However, Figure 2a in 
Nichol {\it et al.}
clearly shows strong correlation
where $\sigma$ is small and $\pi$ is large (50 - 80 $h^{-1}$Mpc) (This is a strong
indication of selection bias and is not discussed by 
Nichol {\it et al.}). 

Figures 5b-d show isocontours of $\xi(\sigma, \pi)$ for the Northern magnitude-limited,
whole-sky magnitude-limited and whole-sky statistical samples. Notice
that the strong correlations with $\sigma \le 20 h^{-1}$Mpc and $50 \le \pi \le 80 h^{-1}$Mpc
are no longer evident in these $R =ge 1$ cluster samples.
We remind
the reader that only the $R \ge 1$ clusters were defined as a statistically complete
sample by Abell (1958). Unfortunately, until now, there simply were not enough $R \ge 1$ cluster
redshifts to perform such an analysis for a magnitude-limited sample. 
While the strong indication of anisotropy is no longer evident in Figures 5b-d, 
there is a residual anisotropy evident in
the 2:1 elongation of $\xi(\sigma, \pi)$ (where  $\sigma \le 20 h^{-1}$Mpc and
$0 \le \pi \le 40 h^{-1}$Mpc).

For the sake of clarity,
we will perform all further analyses
solely on the whole-sky magnitude-limited sample. This choice is based on the 
large number of clusters in this sample and the similarities in the correlation
function results
between the whole-sky magnitude-limited 
and the whole-sky statistical samples (see Figures 4b,c and 5c,d).

\subsection{The Effects of Elongated Superclusters}

We determined which cluster pairs were causing the excess in $\xi(\sigma, \pi)$ in the region
where  $\sigma \le 10 h^{-1}$Mpc and
$20 \le \pi \le 40 h^{-1}$Mpc. There are 19 pairs within these ranges of $\sigma$ and $\pi$
and we discovered that eight (8) of these were associated with
either the Corona Borealis supercluster or the Ursa Majoris supercluster.
Random
catalogs produced only $\sim 8$ pseudo-cluster pairs within these ranges of  $\sigma$ and $\pi$.
Using
methods similar to that of Jaaniste {\it et al.} (1998), we fit ellipsoids to the clusters within
the two superclusters to determine their shapes and orientations. Ursa Majoris is
comprised of clusters A1291, A1318, A1377, A1383, and A1436. Its ellipsoidal fit produces
the axis ratio 1.0:0.2:0.1, which is highly filamentary. The angle between the semi-major 
axis and the line-of-sight is $14^{\circ}$. We also fit the six major clusters
in Cor Bor (A2061, A2065, A2067, A2079, A2089, and A2092) producing the axis ratio
of 1.0:0.4:0.25 with the angle between the semi-major axis and the line-of-sight at
$22^{\circ}$. We note these results are slightly different from those of
Jaaniste {\it et al.}  due to the fact that we are using only $R \ge 1$ clusters.
While it is obvious that the (excess) pairs with $\sigma \le 10 h^{-1}$Mpc and
$20 \le \pi \le 40 h^{-1}$Mpc are mostly members of superclusters
that are elongated roughly
along the line of sight, it is not so obvious that these pairs are the cause
of the elongation of $\xi(\sigma, \pi)$ in this region.

While we could remove these superclusters from the catalog and re-examine the contour
plots of $\xi(\sigma, \pi)$, we saw in the previous section that this could affect the
amplitude and slope of the standard two-point correlation function, $\xi(r)$. By removing
certain clusters from the catalog we are
altering the inherent structure of the catalog (unless, of course, the clusters
removed are spurious). Instead,
we have chosen to rotate these superclusters until their semi-major axes are perpendicular to the
line-of-sight. We then recalculated the standard two-point correlation function and
found no difference in the amplitude or slope (see Figure 6 and Table 2).  
Figure 7 shows the isocontours of $\xi(\sigma, \pi)$ for the new whole-sky magnitude-limited
sample with the Ursa Majoris and Corona Borealis superclusters rotated so that
their semi-major axes are nearly perpendicular to the line-of-sight. The region
of strong clustering shows no elongation whatsoever, and there are only very weak
correlations beyond $25h^{-1}$Mpc.

At this point, we need to qualify the reality of these two superclusters, Ursa Majoris and Corona
Borealis. All of the clusters within these superclusters have multiple observed galaxy redshifts, although
Corona Borealis has been studied in greater detail (Schuch, 1981, 1983; Postman {\it et al.} 1986, 1988;
Small {\it et al.} 1997a,b, 1998). PHG, whose dataset included both the Ursa Majoris and
the Corona Borealis superclusters, looked for contamination within their sample by
recreating the spatial distribution of $R \ge 0$ clusters using realistic cluster profiles. They found only
five $R \ge 1$ clusters that show contamination, one of which was A2061 within Corona Borealis.
When A2061 is excluded from our samples, we find no difference in any of our results. According to
PHG,  none of the 
clusters in Ursa Majoris show contamination.	
While the MX Survey has added $\sim 70$ new northern clusters in the magnitude extension to
m$_{10} = 16.8$,
none of these clusters are 
near enough to Ursa Majoris or Corona Borealis to cause new possible contaminations related
to those superclusters.

Since projection contamination has been suggested as a reason for
dismissing the Abell catalog for large-scale structure studies,
it is worthwhile examining even the most unlikely 
case where the Ursa Majoris supercluster contains mostly
$R =0$ clusters (or poor groups) suffering from extreme projection (thus inflating
apparent richnesses). In this
analysis,  we find no change in the
correlation length or slope (see Table 2) after excluding four of the five
Ursa Majoris cluster members from the catalog. We can take this analysis one step
further and exclude {\it any} Abell/ACO clusters that show projection problems. In this case,
we looked at the galaxy velocity distributions for clusters published in
Zabludoff {\it et al.} (1993), Mazure {\it et al.} (1996), and Slinglend {\it et al.}  (1998).
We reduced the richness classification by one level whenever the total number of galaxies within
any cluster was less than 60\% of the total number of galaxies observed in the cluster field.
(Note: some of the ACO clusters had already been excluded by the $N_{gal} \ge 54$ cut-off.).
In the end, we excluded 19 clusters from the original 289 in the whole-sky magnitude-limited sample
(an additional five clusters were reduced from $ R = 2$ to $R = 1$).
After re-examining the two-point correlation function, we find almost no change in the
correlation length or slope (see Table 2). 

The reader may also be concerned with the fact that we have detected two elongated superclusters along
the line-of-sight within a catalog that contains a small number of superclusters. To examine
the probability of such an event, we performed
percolation analyses on the whole-sky magnitude-limited sample at two percolation lengths,
$b = 20 h^{-1}$Mpc
and $b = 30 h^{-1}$Mpc,
corresponding to spatial density enhancements ($n/\bar{n}$) of $\sim 20$ and $\sim 5$.
At $b = 20 h^{-1}$Mpc we find seven superclusters with five or more cluster members and four
of these superclusters show elongation similar to or greater than that of Corona Borealis.
To see how often we would expect one of these elongated superclusters to lie
within $25^{\circ}$ of the line-of-sight (recall, the semi-major axis of Corona Borealis is 
$22^{\circ}$ off the line-of-sight) we perform a geometrical calculation among
randomly oriented structures.
The probability of finding a single elongated supercluster along the line-of-sight is
equal to the solid angle of a circle with diameter 25$^{\circ}$ divided by $2\pi$
steradians or 
$\sim 0.10$. The probability of finding one of the four elongated superclusters
along the line-of-sight is $\sim 0.30$. There is a 10\% chance of finding two of these four
superclusters to lie within $25^{\circ}$ of the line-of-sight.
At $b = 30 h^{-1}$Mpc we find eight superclusters with five or more cluster members and five
of these superclusters show elongation similar to or greater than that of Corona Borealis.
In this case, there is a 15\% chance of finding two of the five elongated superclusters
to lie within 25$^{\circ}$ of the line-of-sight. These results indicate that if four (or five) elongated
superclusters were randomly oriented in space, a central observer would have reasonable 
chance,
10\% - 15\%, 
of finding two superclusters within $25^{\circ}$ of the line-of-sight.
Finding two such superclusters in the Abell/ACO catalogs is  modestly
probable.
We refer the reader to Batuski {\it et al.} (1999) for a more detailed
examination of filamentation in the Abell/ACO catalogs.

From this analysis, we can conclude that two evidently real
structures, elongated along the line-of-sight, are
the cause of the previously detected anisotropy within the Abell/ACO catalogs. The 2:1 elongation
of $\xi(\sigma, \pi)$ seen in Figures 5b-d would not appear in these contour plots,
had the two superclusters been oriented perpendicular to the line-of-sight.
Thus, the anisotropy present in Figures 5b-d is the result of superclustering
along the line of sight,
not an indicator of observational bias or high peculiar velocities. In addition, the
fact that we can remove this anisotropy from the catalog (via supercluster rotation)
{\it without} changing $r_0$ calls the previously
mentioned `corrective' techniques into question.

Figures 5 and 7 indicate the level of anisotropy in a visual, somewhat qualitative way.
Attempts have been made to quantify this anisotropy by convolving the derived
form of the correlation function with a Gaussian of chosen width
(Peacock \& West 1992; Estathiou {\it et al.} 1992; Dalton {\it et al.} 1994): 
\begin{equation}
\xi(\sigma,\pi) = \frac{r_0^{-\gamma}}{\sqrt{2\pi}\sigma_v}\int_{-\infty}^{\infty} 
[\sigma^2 + (\pi-x)^2]^{\gamma/2}e^{\frac{-x^2}{2{\sigma_v}^2}}dx.
\end{equation}
This attempts to match the observed $\xi(\sigma,\pi)$ with the $\xi(r)$ obtained
from the best power law fit.
We must also note that an assumed
Gaussian distribution for $\xi$ is most likely inappropriate because of $\xi$'s 
power-law form. This is not an attempt to determine cluster peculiar velocities. We
are merely providing these results for comparison to other such analyses.

For this analysis, we can look only at a region constrained by $0 \le \sigma \le 15$
($h^{-1}$Mpc) 
where the number of paircounts is appreciable.
Figure 8a shows  $\xi(\sigma, \pi)$ plotted against $\pi$ for cluster pairs
with projected separations in the range $0 \le \sigma  \le 15 (h^{-1}$Mpc)  and
total separations $ r \le 100 h^{-1}$Mpc. The solid histogram is the PHG statistical
sample while the dash-dot histogram is the whole-sky magnitude-limited sample.
The smooth lines are the respective histograms convolved with a Gaussian as in Equation (11)  with
$\sigma_v = 700$ km s$^{-1}$ and the appropriate slope and correlation length.
As an indicator of anisotropy within a cluster sample, we measure the mean value
of $\xi(\sigma,\pi)$ for $\sigma \le 15h^{-1}$Mpc and $\pi \ge 20h^{-1}$Mpc
where we expect $\xi(\sigma,\pi)$ to be small.
The mean values of the correlation function beyond $\pi = 20 h^{-1}$Mpc are
$\bar{\xi}(\sigma,\pi) = 0.53 \pm{0.56} $ for the PHG sample and 
$\bar{\xi}(\sigma,\pi) = 0.32 \pm{0.60} $ for the whole-sky sample. Notice
the large reduction in $\xi(\sigma, \pi)$ beyond $40 h^{-1}$Mpc between the PHG
sample and the whole-sky sample. Figure 8b compares
the whole-sky sample before and after the Ursa Majoris and Cor Bor supercluster
rotations. Again, the curves are the respective histogram
convolved with a Gaussian.  
Notice the drop in correlation in the range $20 \le \pi \le 50 h^{-1}$Mpc. This is consistent
with the drop in correlations noticeable in Figures 5c and 7.
The mean value of the correlation function beyond 20 $h^{-1}$Mpc for  the
rotated case is $\bar{\xi}(\sigma,\pi) = 0.16 \pm{0.49} $. While the PHG case shows
some strong correlations out to $70 h^{-1}$Mpc, the new $R \ge 1$ whole-sky magnitude-limited sample
has essentially no detectable correlations beyond $40 h^{-1}$ Mpc. After 
rotating the Ursa Majoris and Corona Borealis superclusters so that they are not elongated
along the line of sight, we find no detectable correlations beyond $20 h^{-1}$ Mpc. This is comparable
to the claims made by Efstathiou {\it et al.} (1992) that the APM cluster catalog shows no
line-of-sight correlations
beyond $25 h^{-1}$Mpc. An analysis of Figure 3 in Efstathiou {\it et al.} (1992) shows 
$\bar{\xi}_{APM}(\sigma,\pi) \sim 0.10$ beyond $20 h^{-1}$Mpc, which is very similar to what we
find in the Abell/ACO catalogs
after rotating the two superclusters. 

\section{Conclusion}

The results presented in this work provide strong evidence for the
{\it lack} of observational selection bias and projection effects in the $R \ge 1$ sample
of Abell clusters. While some earlier studies indicated that anisotropies and
spurious cluster selection greatly affected correlation analyses, 
those results were based on smaller and/or incomplete samples,
samples with a large number of single-redshift cluster distances,
and samples containing $R = 0$ clusters. The MX Survey has provided a
large, complete dataset of multiple-redshift $R \ge 1$ Abell cluster distances
for analyses of large-scale structure in the local universe.

We have calculated the mean space
density of $R \ge 1$ clusters to be
$\bar{n}_p = 6.6 (\pm 0.6) \times 10^{-6} h^{3}$Mpc$^{-3}$
for the northern Abell clusters and $ \bar{n}_p = 7.6 (\pm 0.9) \times 10^{-6}$
$h^{3}$Mpc$^{-3}$ for the southern ACO clusters. 
This result supports previous claims 
that the ACO catalog includes slightly poorer $R \ge 1$ clusters than the northern Abell catalog.
Power-law fits of the two-point spatial correlation function of the MX samples
are presented in Table 2 and in Figure 4.  We find $19.4 \le r_0 \le 23.4 h^{-1}$Mpc,
$-1.92 \le \gamma \le -1.83$
for the different subsets of $R \ge 1$ clusters examined. 

We have used the largest rich cluster data set available to
date to look for line-of-sight anisotropies within the Abell/ACO catalogs.
An examination of the correlation function
separated into its line-of-sight ($\pi$) and perpendicular-to-line-of-sight ($\sigma$)
components show that the strong
anisotropy present in the PHG sample (where $\xi(\sigma,\pi) > 1$ with 
$\sigma \le 20h^{-1}$Mpc and $50 \le \pi \le 80h^{-1}$Mpc in Figure 5a)
is not present in our samples  (Figures 5b-d). 
The remaining anisotropies in our samples, indicated by the 2:1 elongation of
$\xi(\sigma, \pi)$ (where $\sigma \le  20h^{-1}$Mpc and $\pi \le 40h^{-1}$Mpc),
can be explained by the orientation of two elongated superclusters, Ursa Majoris
and Corona Borealis. After rotating these well-established superclusters so that their semi-major
axes are perpendicular to the line-of-sight, we find no indication of
anisotropy in the contour plots of $\xi(\sigma,\pi)$ (see Figure 7).
Therefore, the 2:1 elongation of $\xi(\sigma, \pi)$ can be considered
a true characteristic of the sample examined, not an indicator of line-of-sight
selection biases.

We performed percolation analyses on the whole-sky magnitude-limited sample and
find 4-5 elongated superclusters with axes ratio of 2.5:1:1 or greater
(depending on the choice of
percolation length). If these elongated superclusters were distributed in space
with random orientations, the probability of finding two
within 25$^{\circ}$ of the line-of-sight
is 10\% - 15\%, which is not unduly unlikely. 
Thus, it is not surprising to find anisotropies (as detected by contour plots
of the correlation function)
as the result of superclustering along the line-of-sight.

The correlation length and slope of the two-point spatial correlation
function, $\xi(r)$, are very robust for the whole-sky magnitude-limited sample.
After rotating Ursa Majoris and Corona Borealis so that they are
perpendicular to the line-of-sight,
both $r_0$ and $\gamma$ remain unchanged. Since these rotations
remove any evidence of line-of-sight anisotropies within the catalog, we
conclude that
$r_0$ is not inflated (compared to the APM clusters) by these anisotropies. 
We also find no change in $r_0$ and $\gamma$ after excluding 19 clusters
(from various sources) that may have heavy foreground/background contamination.
This work conflicts with earlier conclusions by Sutherland (1988) (whose results were then expanded upon 
in Efstathiou {\it et al.} (1992) and Nichol {\it et al.} (1992)), that the Abell catalog suffers from
a large amount of
spurious cluster selection which, in turn, inflates $r_0$.
While there are certainly individual
clusters that show heavy background/foreground contamination, they make up only 10\% of
the total in our magnitude-complete samples and their exclusion has no effect
on the amplitude or slope of the correlation function.
If, however, we remove actual structures from the dataset, such as
the Corona Borealis and Microscopium superclusters,  $r_0$
decreases (by up to 20\%) and $\gamma$ steepens (from $\sim 1.8$ to $ 2.1h^{-1}$Mpc).
The high value of $r_0$ for $R \ge 1$ Abell clusters continues to rule out
the standard CDM models and future simulations must be able to reproduce
large-scale clustering out
to scales of $\sim 50 h^{-1}$Mpc.

\acknowledgments
We wish to thank Adrian Melott for helpful discussions on the manuscript. 

KAS, DJB and CM were
supported in this work by National Science Foundation Grant AST-9224350.

CM was funded in part by the National Aeronautics and Space Administration
and the Maine Science and Technology Foundation 
 
This research has made use of the NASA/IPAC Extragalactic Database (NED),
which is operated by the Jet Propulsion Laboratory, California Institute of
Technology, under contract with the National Aeronautics and Space
Administration.

\begin{figure}
\plotone{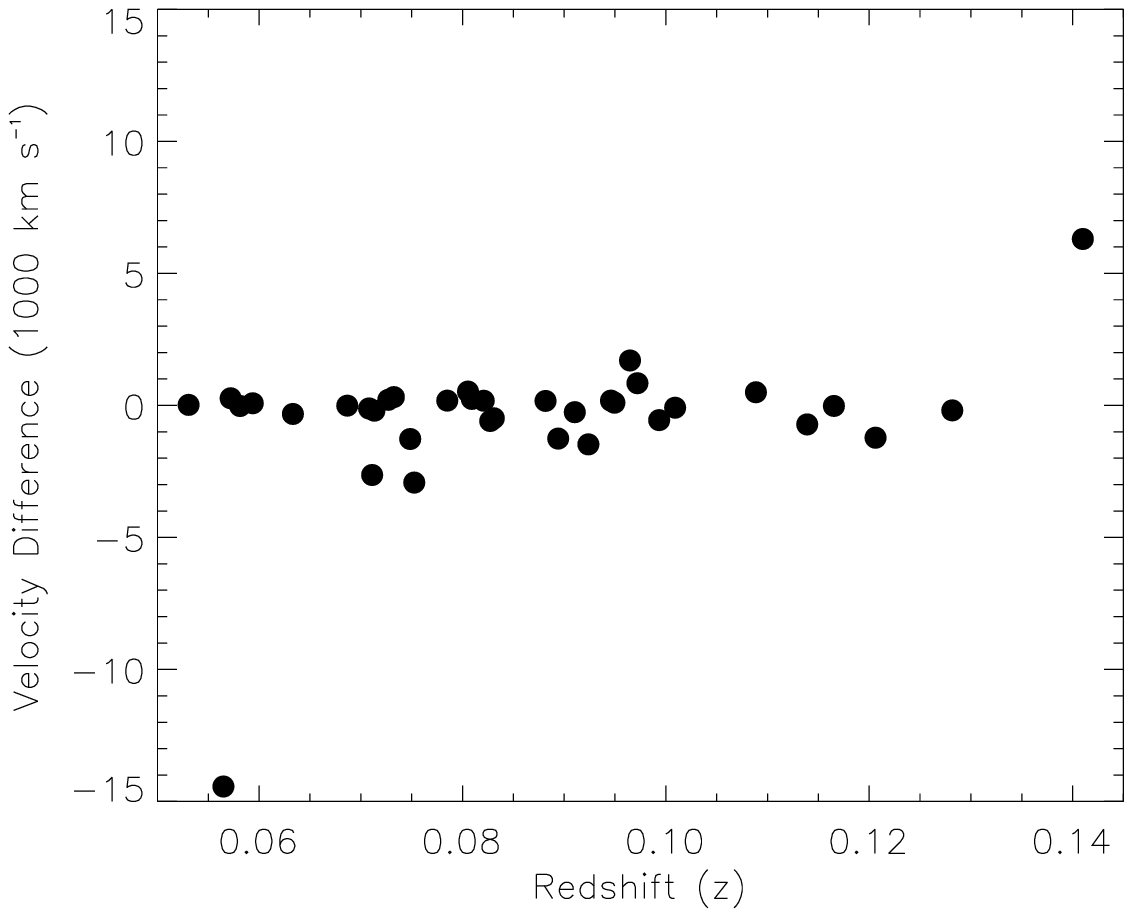}
\caption[]{The differences for cluster velocities
as determined with one measured redshift, compared to velocities
determined with an average of nine galaxy redshifts from the MX Survey.}
\end{figure}
 
\newpage
\begin{figure}
\epsscale{.9}
\plotone{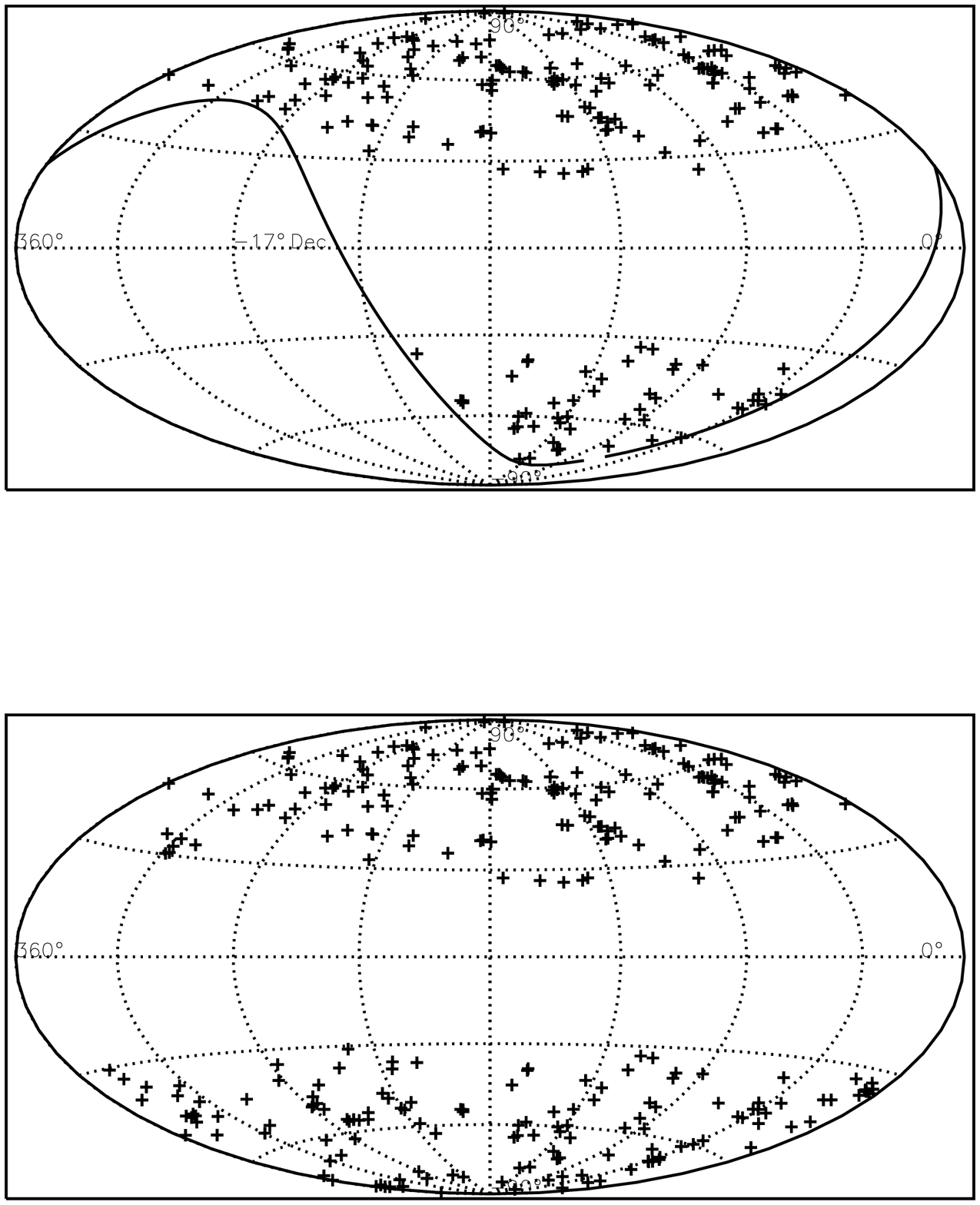}
\caption[]{{\bf Top:} aitoff projection for the Northern magnitude
limited sample of 198 clusters. {\bf Bottom:}  Wholesky magnitude
limited sample of 198 northern clusters and 91 southern clusters.}
\end{figure}

\begin{figure}[ht]
\figurenum{3a}
\plotone{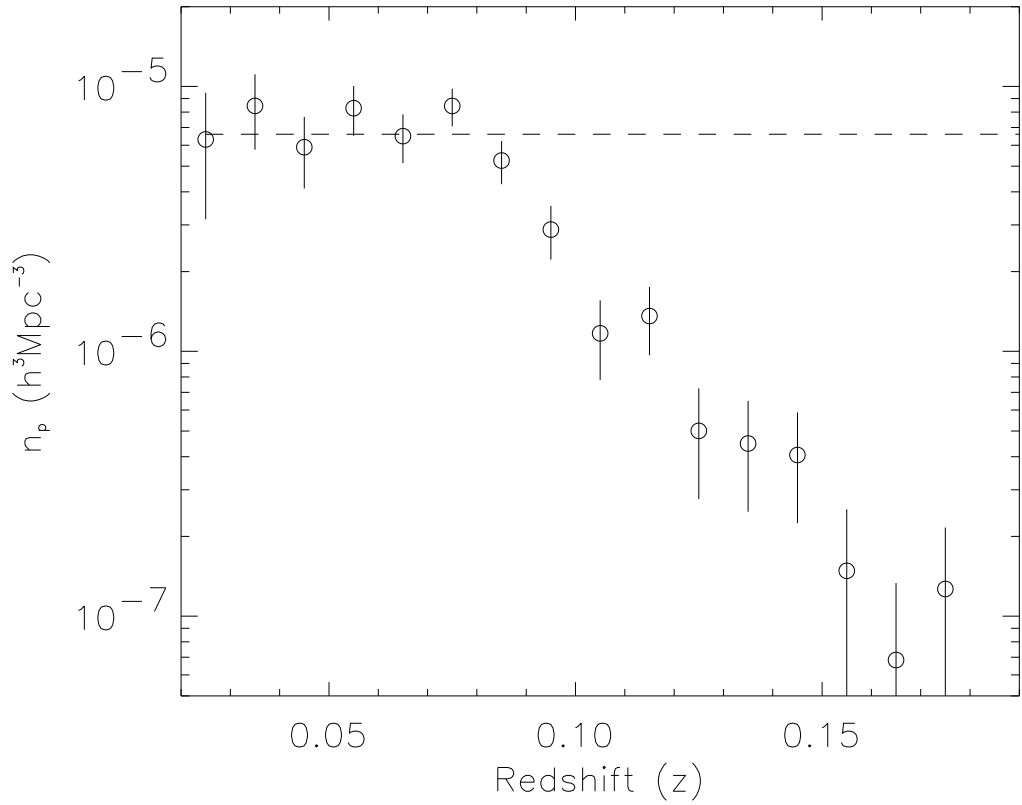}
\caption[]{The  number density (n$_p$)
for the MX Survey $m_{10} \le 16.8$
limited sample of $R \ge 1$ clusters.
The dashed line is the mean density of $6.60\times10^{-6} h^{3}$Mpc$^{-3}$.}
\end{figure}
 
\begin{figure}[ht]
\figurenum{3b}
\plotone{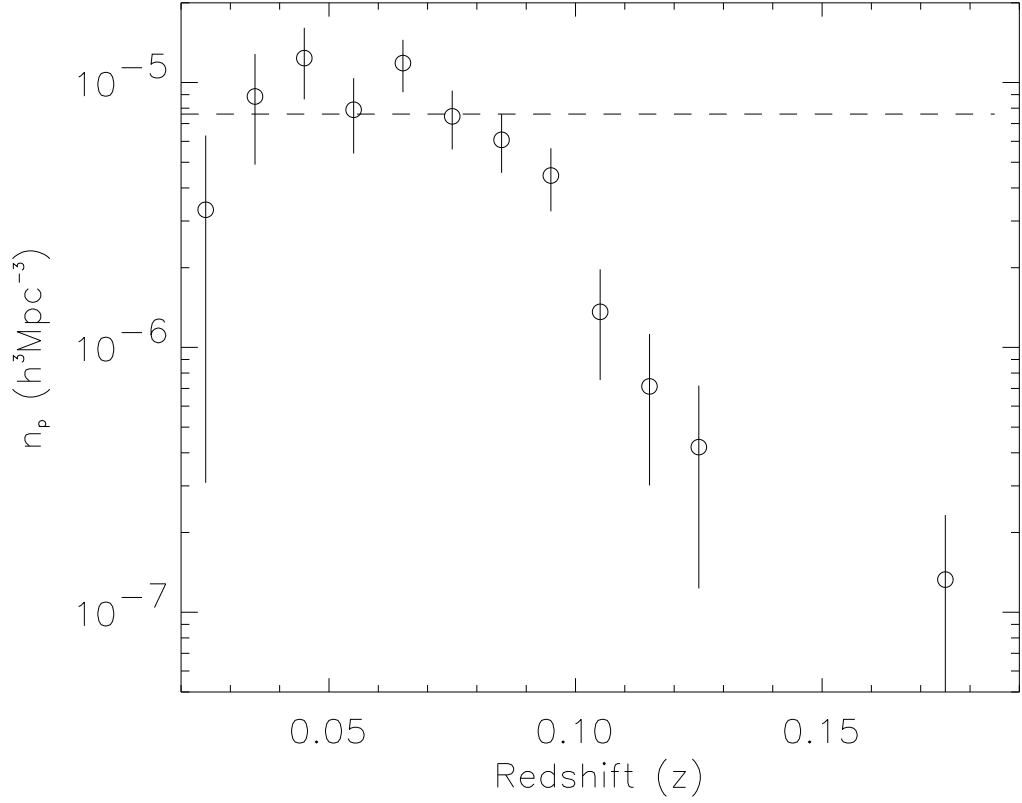}
\caption[]{
The  number density (n$_p$) for
Southern clusters with  m$_{10} \le 16.8$
and $R \ge 1$. The dashed line
is the mean density of $7.60\times10^{-6} h^{3}$Mpc$^{-3}$.}
\vspace{2.0in}
\end{figure}

\newpage
\begin{figure}[ht]
\plottwo{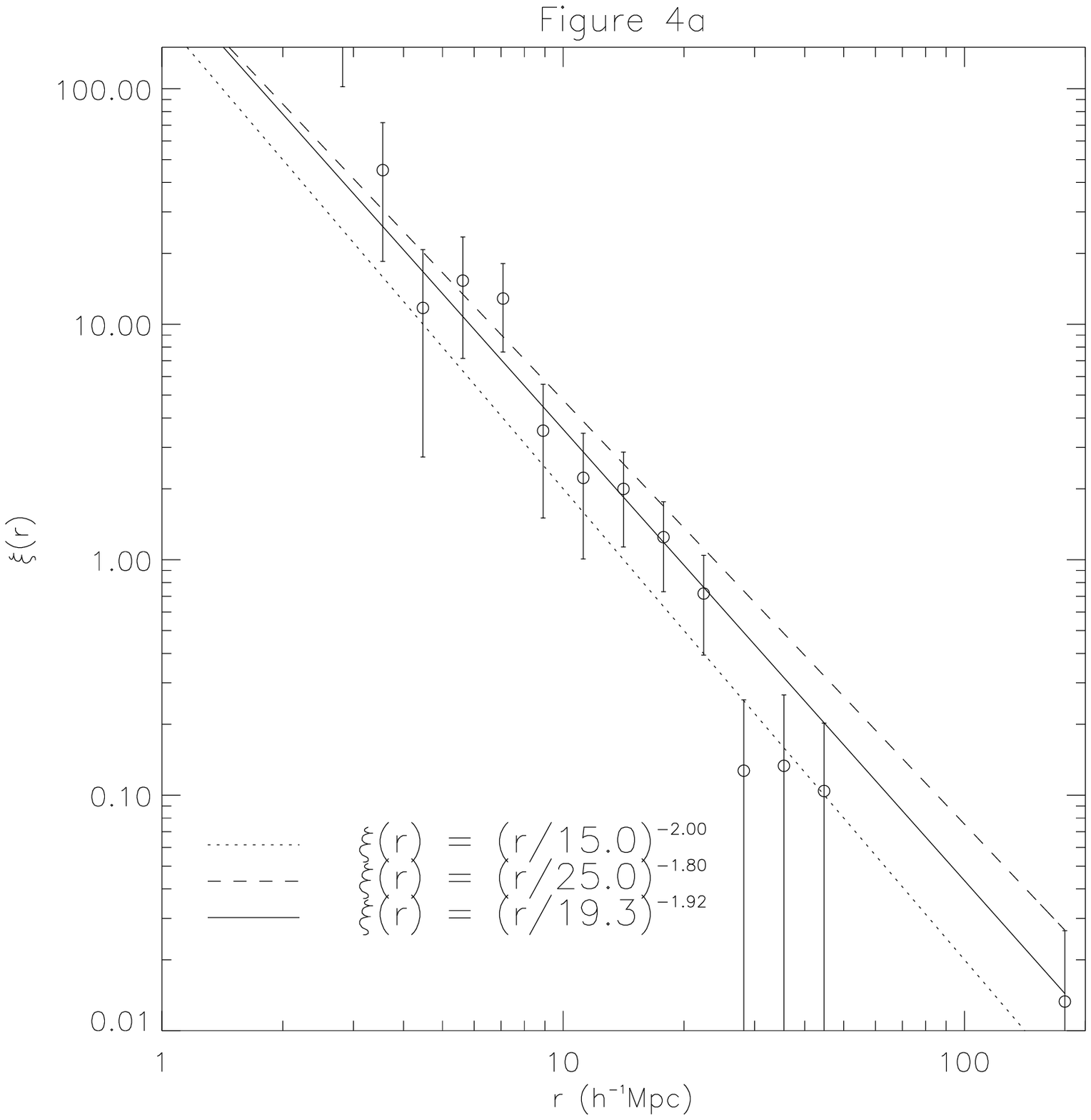}{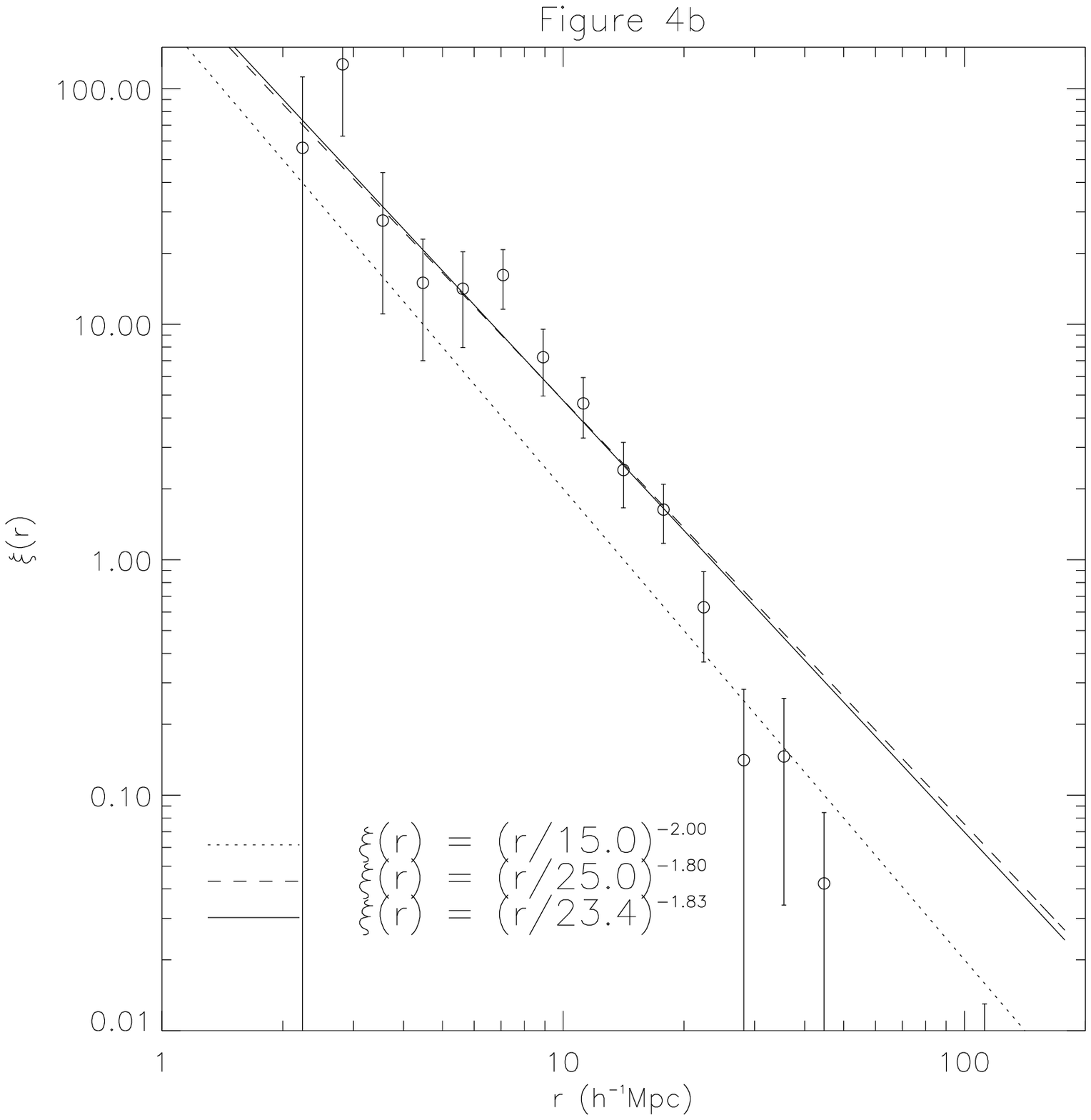}
\end{figure}
\begin{figure}[hb]
\figurenum{4}
\plottwo{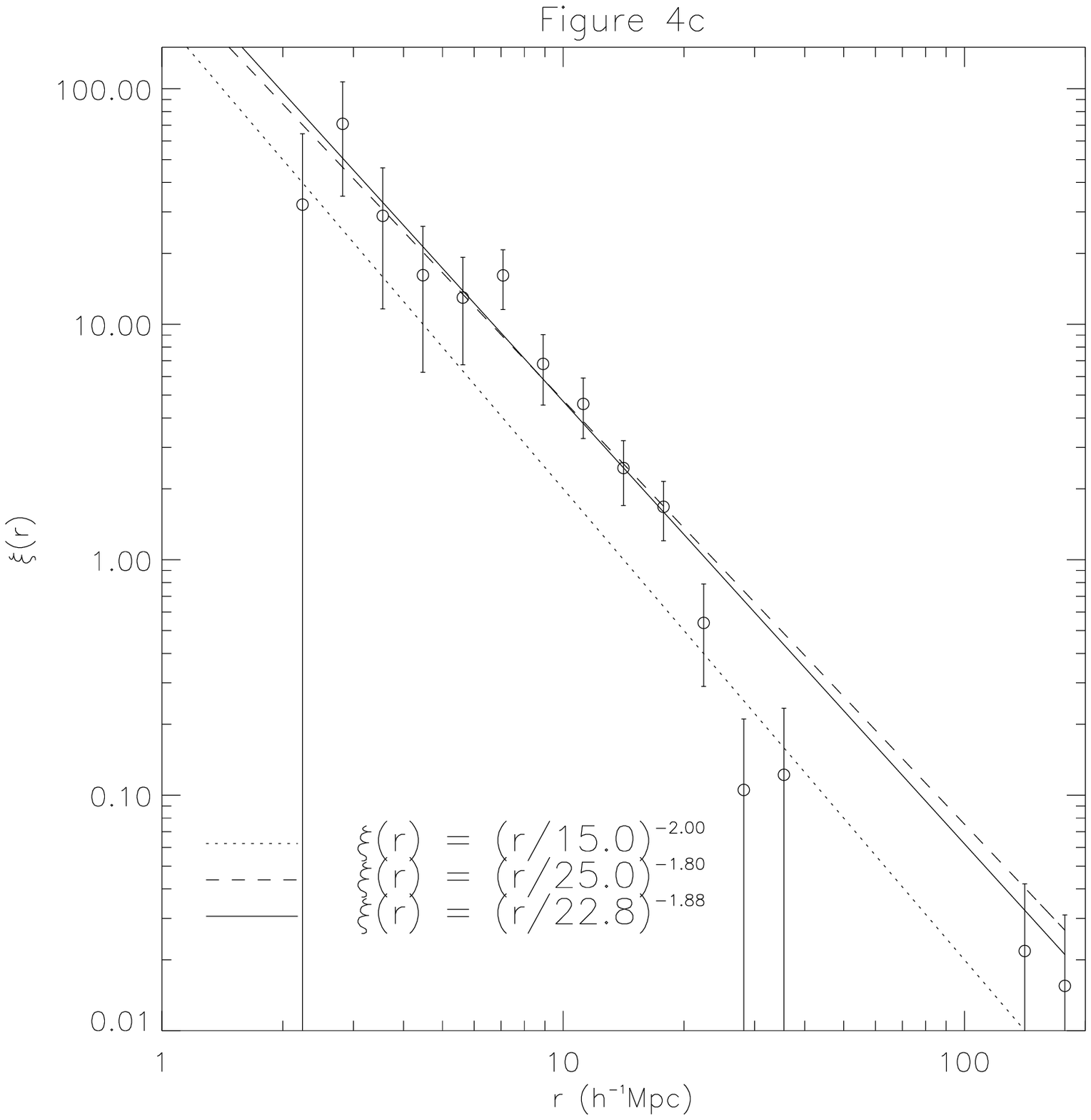}{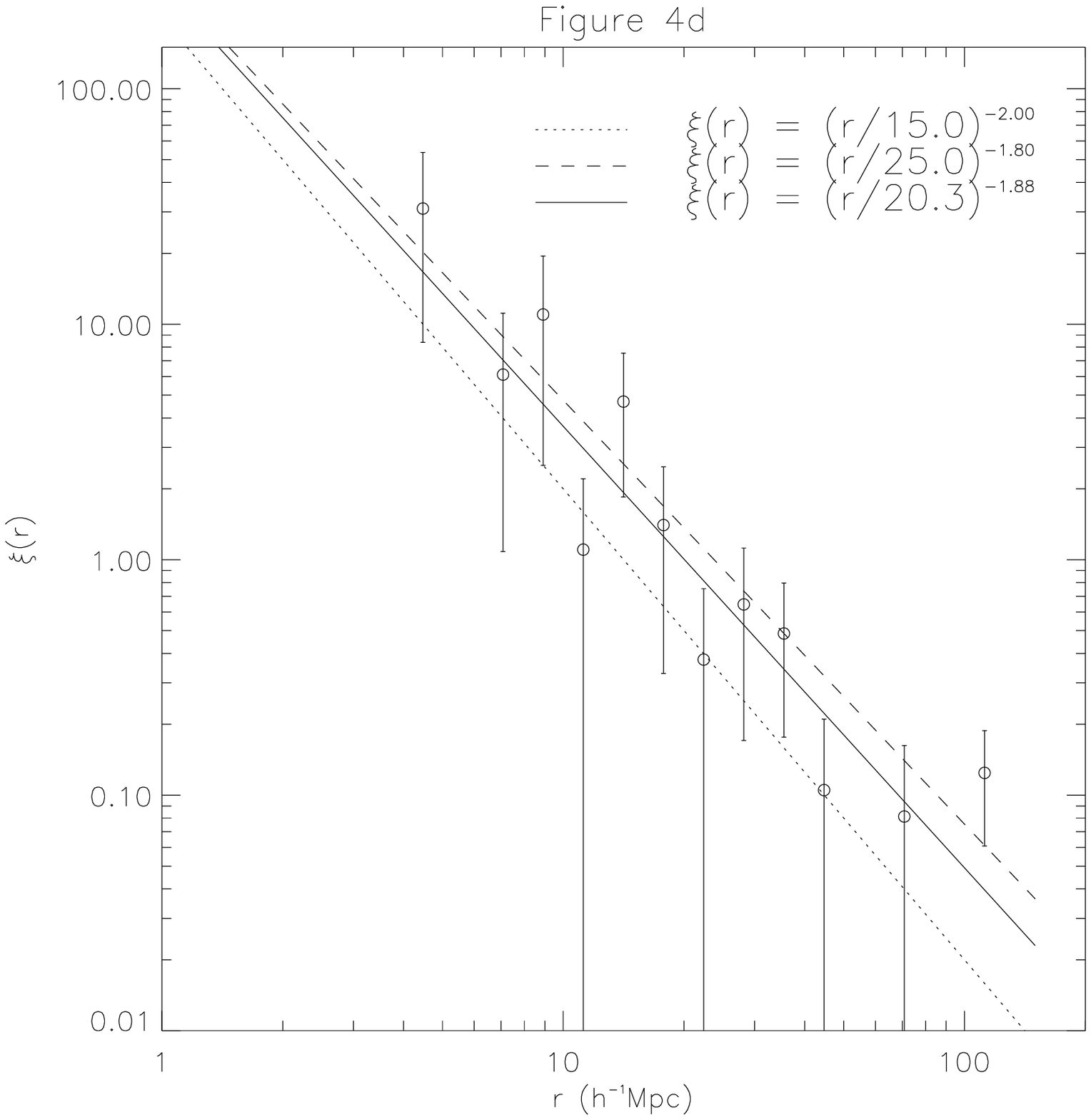}
\caption[]{{\bf Fig 4a:} The northern magnitude-limited sample of 198 Abell clusters with
m$_{10} \le 16.8$ and $R \ge 1$.
{\bf Fig 4b:} The whole-sky magnitude-limited sample of 198 Abell clusters
and 91 ACO clusters with m$_{10} \le 16.8$ and $R \ge 1$.
{\bf Fig 4c:} The whole-sky magnitude and volume-limited sample of
158 Abell clusters and 82 ACO clusters with
m$_{10} \le 16.8$, $R \ge 1$ within $  286 h^{-1}$Mpc ($z = 0.10$).
{\bf Fig 4d:} The cD sample of 104 Abell clusters within $ 350 h^{-1}$Mpc.}
\end{figure}

\pagebreak

\begin{figure}
\figurenum{5}
\plottwo{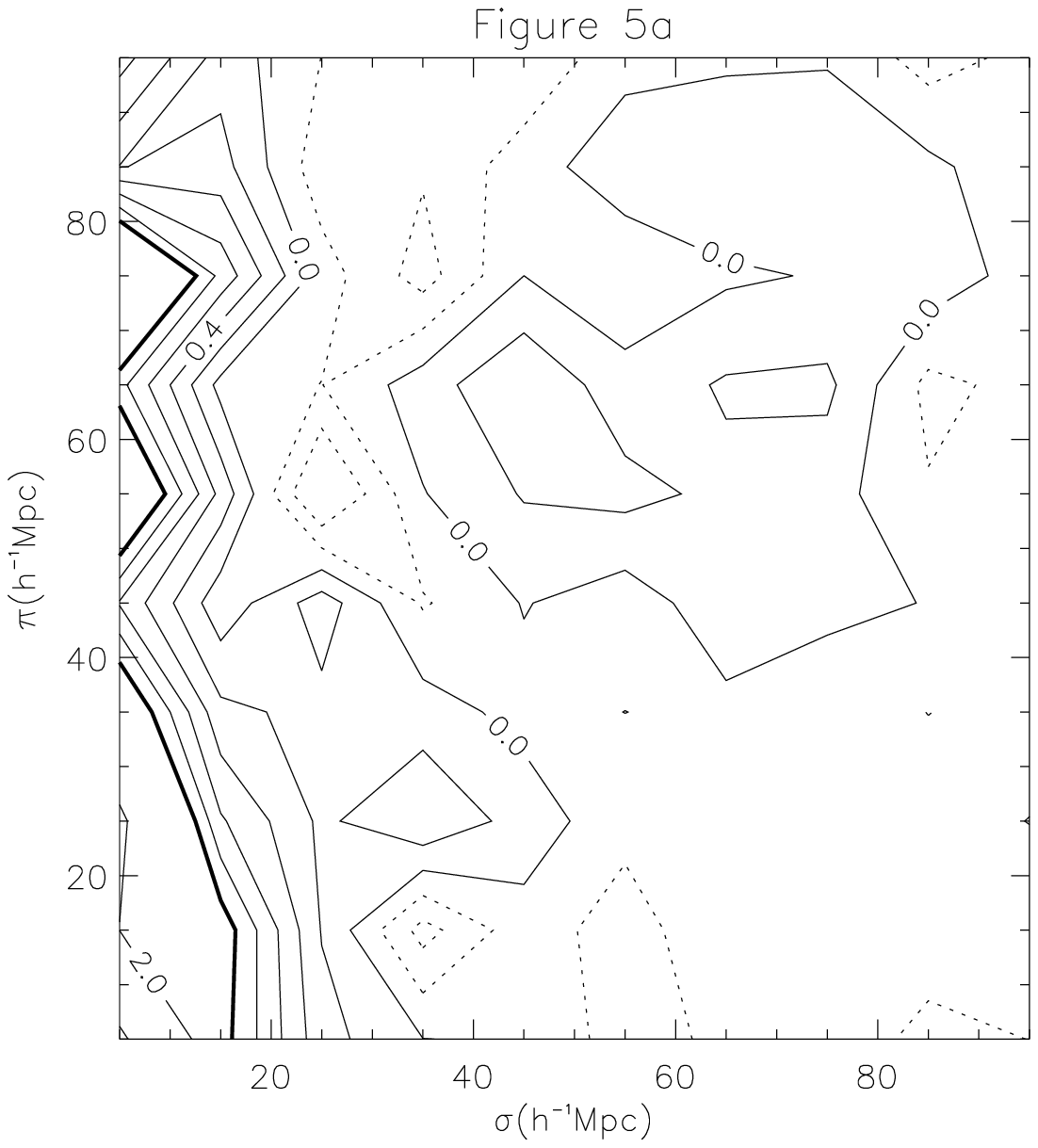}{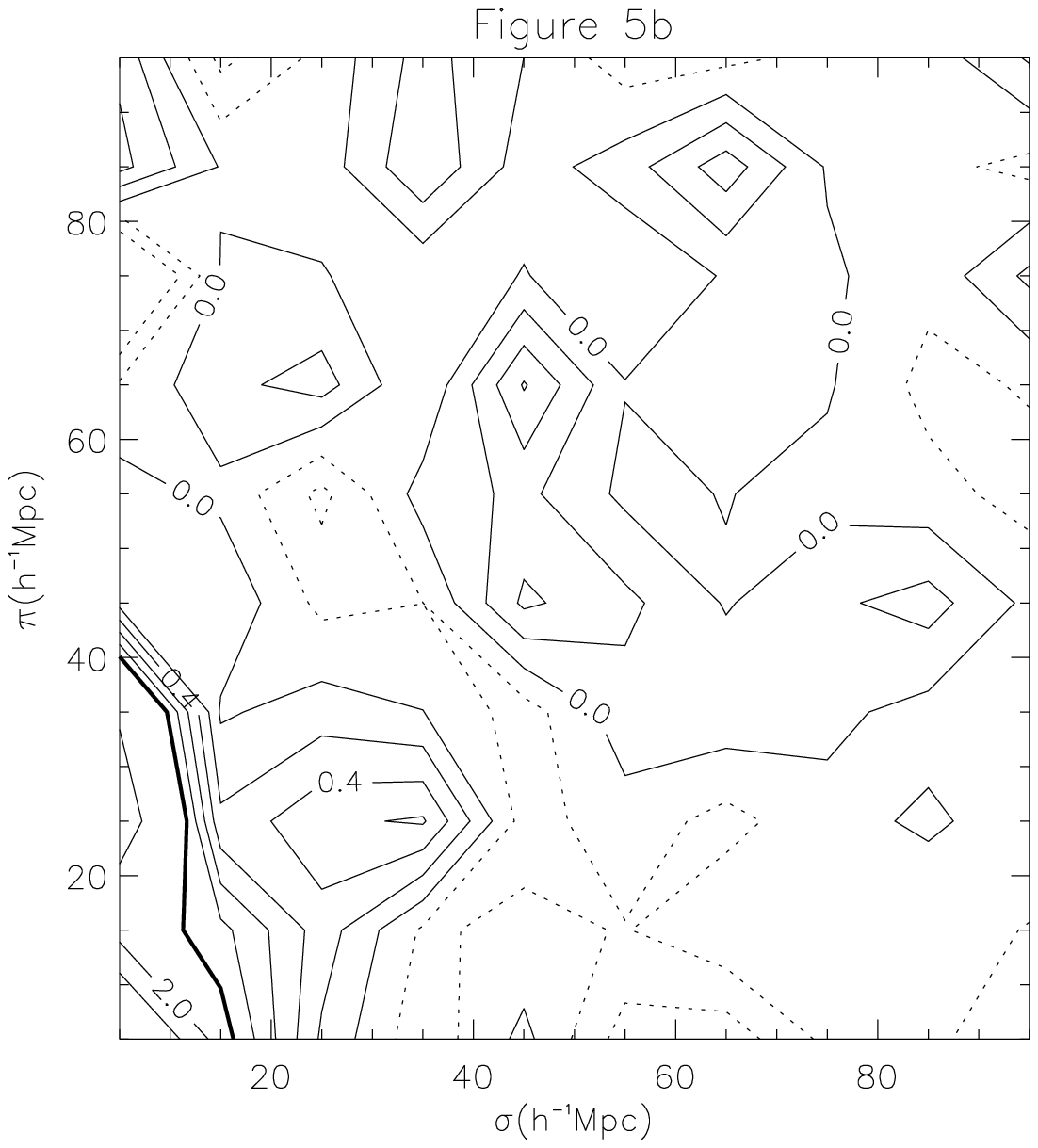}
\end{figure}
\begin{figure}
\figurenum{5}
\plottwo{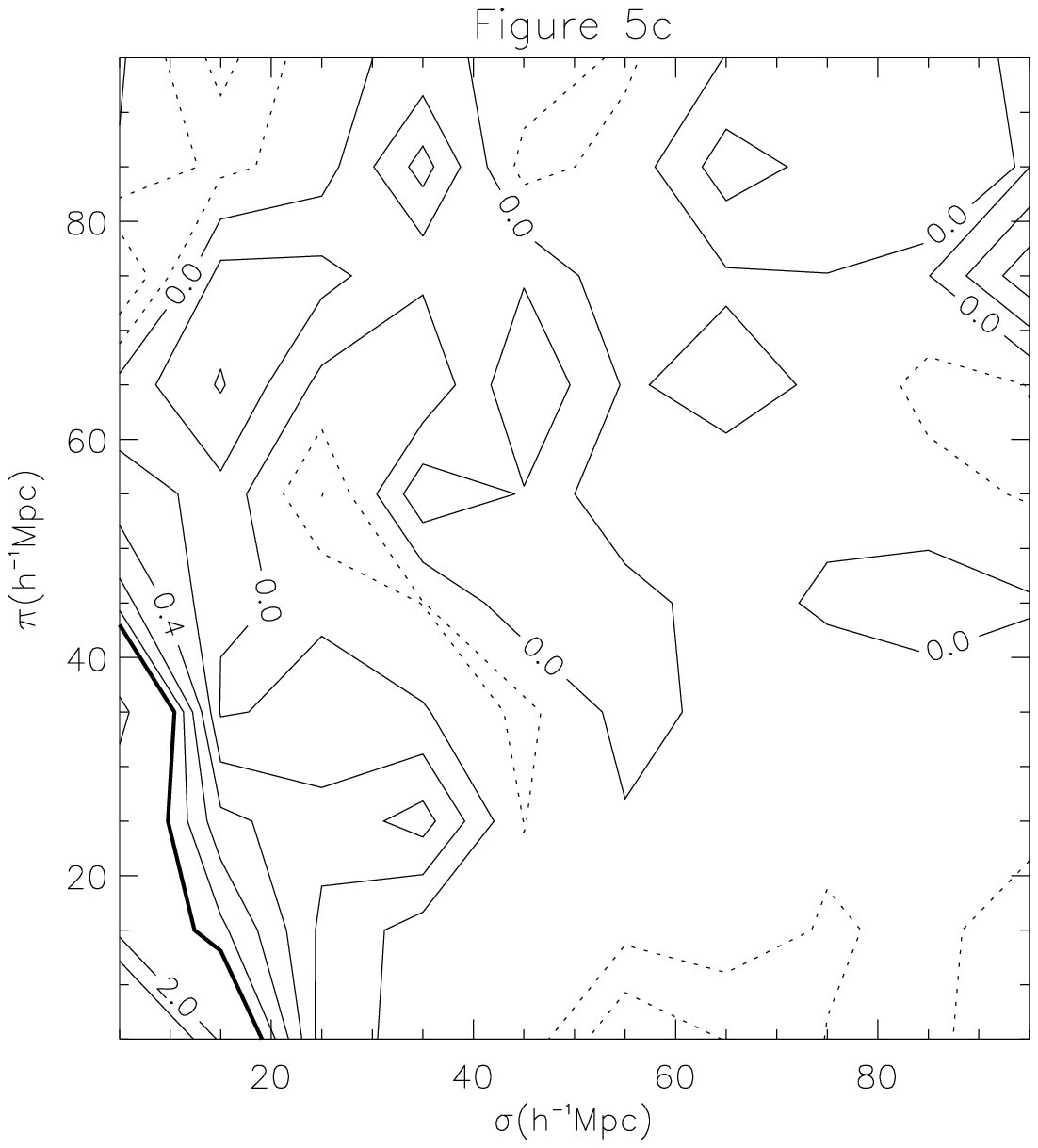}{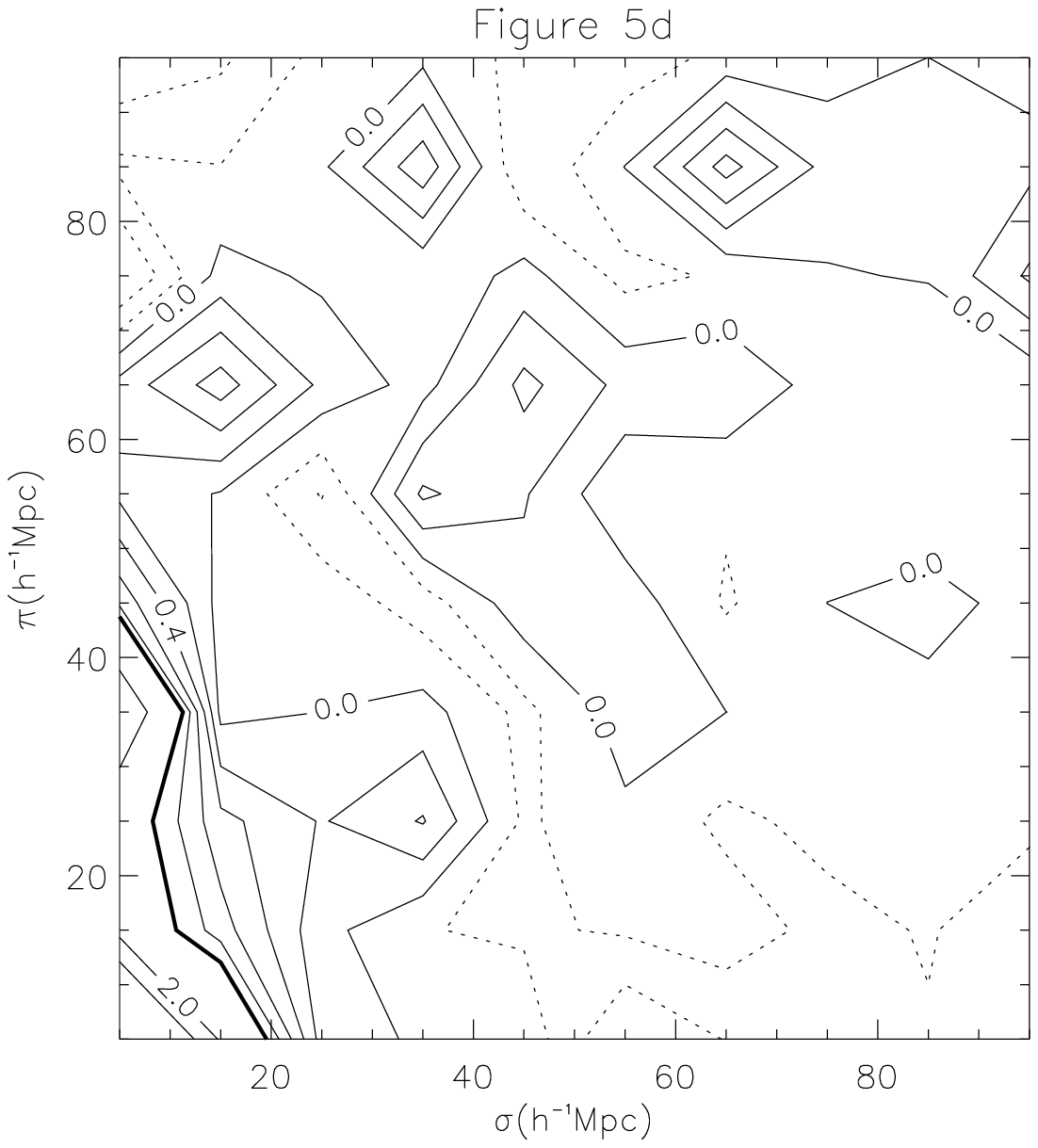}
\caption[] {{\bf Fig 5a:} The contour plot of
$\xi(\sigma,\pi)$ for the original PHG statistical sample of
$R \ge 0$, m$_{10} \le 16.5$, $|b| \le 30$ Abell clusters.
{\bf Fig 5b:} The contour plot of $\xi$ for the
MX northern magnitude-limited
sample of $R \ge 1$, m$_{10} \le 16.8$,
$|b| \le 30$ Abell clusters.
{\bf Fig 5c:} The contour plot of $\xi$ for the
whole-sky magnitude-limited sample of $R \ge 1$, m$_{10} \le 16.8$, Abell/ACO clusters.
{\bf Fig 5d:} The contour plot of $\xi$ for the
whole-sky statistical 
sample.}
\end{figure}

\begin{figure}
\plotone{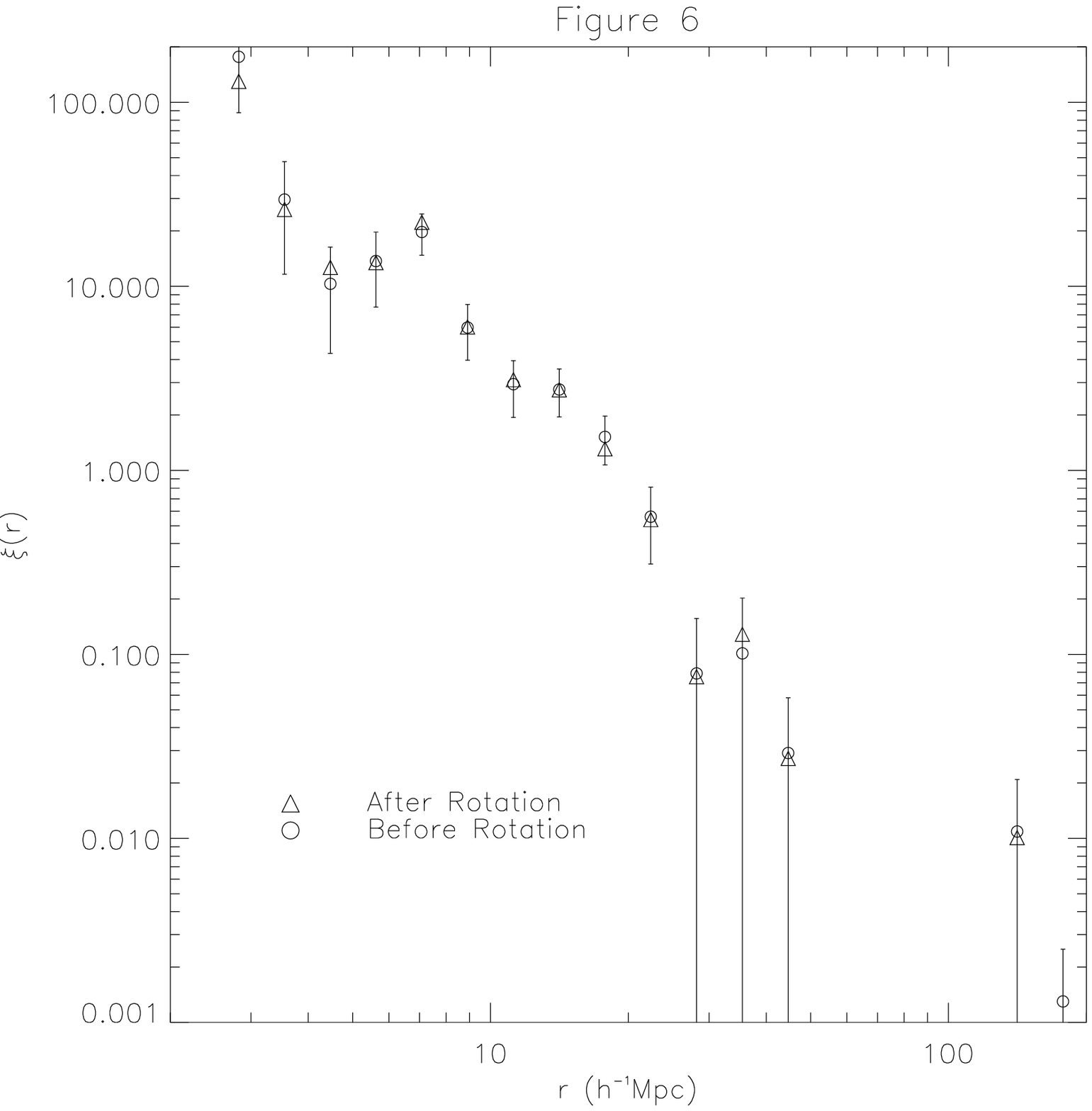}
\figurenum{6}
\caption[]{
$\xi(r)$ for the whole-sky magnitude-limited sample before and
after rotating the Ursa Majoris and Corona Borealis superclusters. 
}
\end{figure}
 
\begin{figure}
\figurenum{7}
\plotone{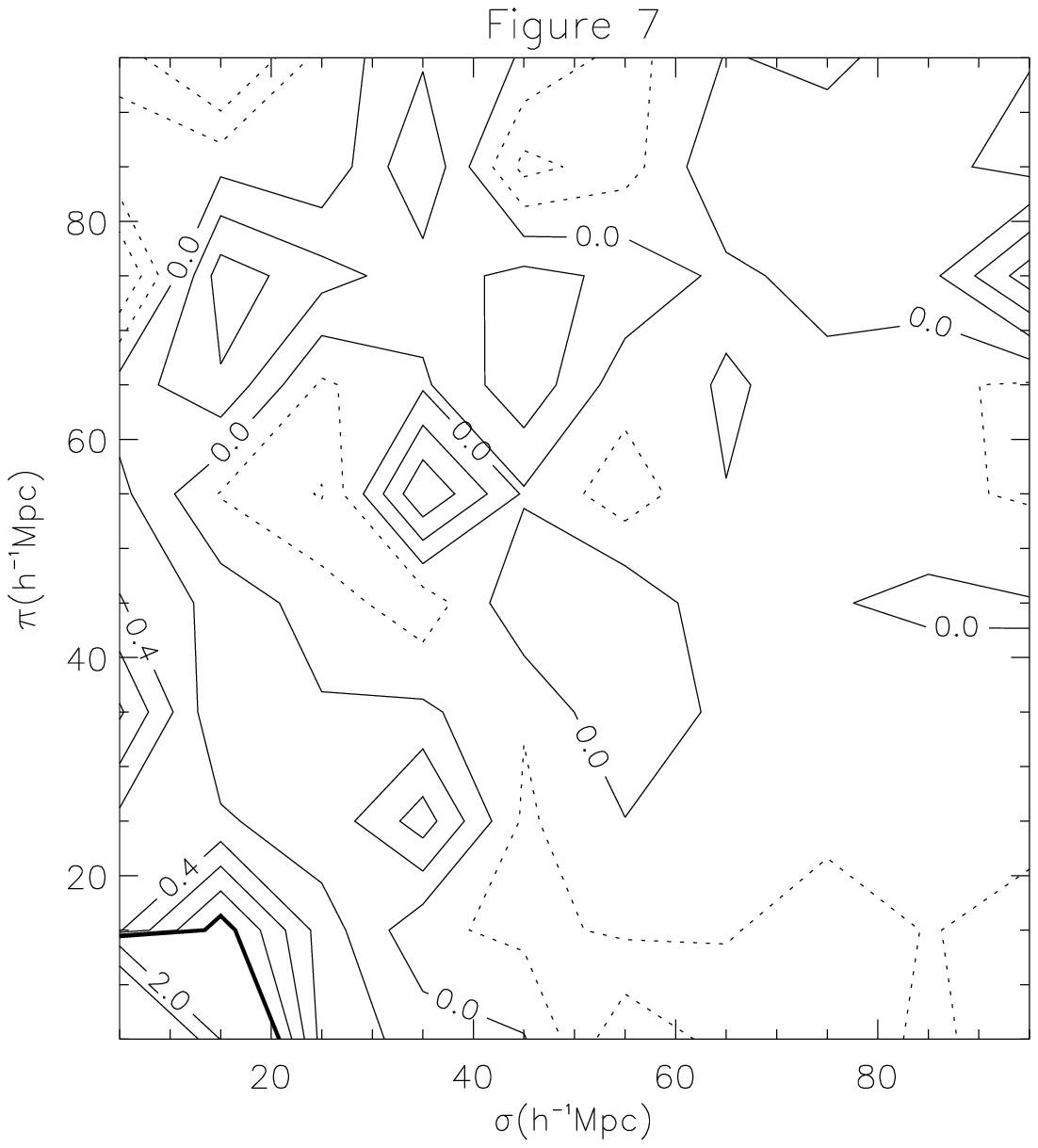}
\caption[]{
$\xi(\sigma, \pi)$ in $10 h^{-1}$Mpc  bins  for the whole-sky magnitude-limited sample
after rotating the Ursa Majoris and Corona Borealis superclusters.
Compare to Figure 5c.
}
\end{figure}

\begin{figure}
\figurenum{8}
\plottwo{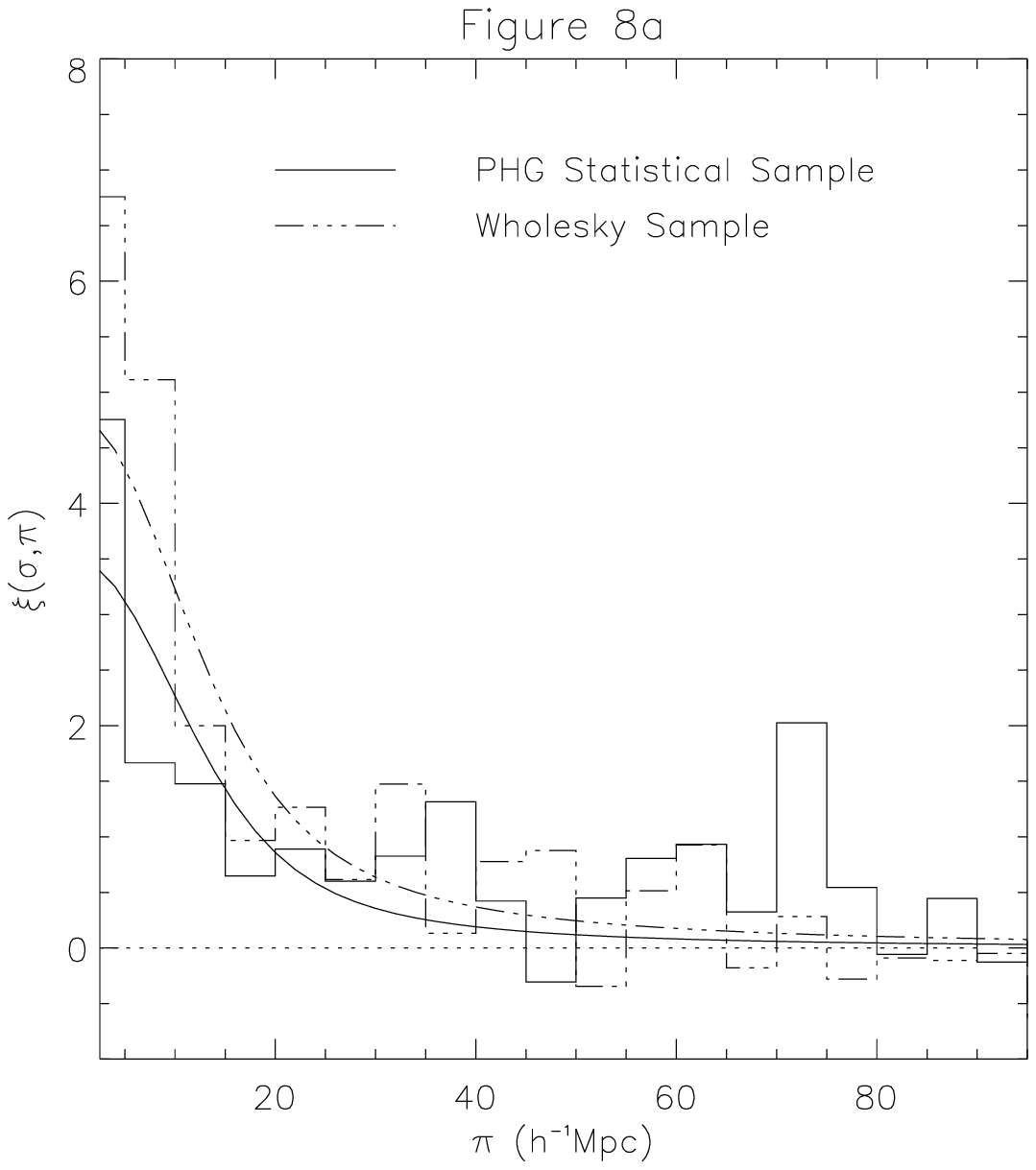}{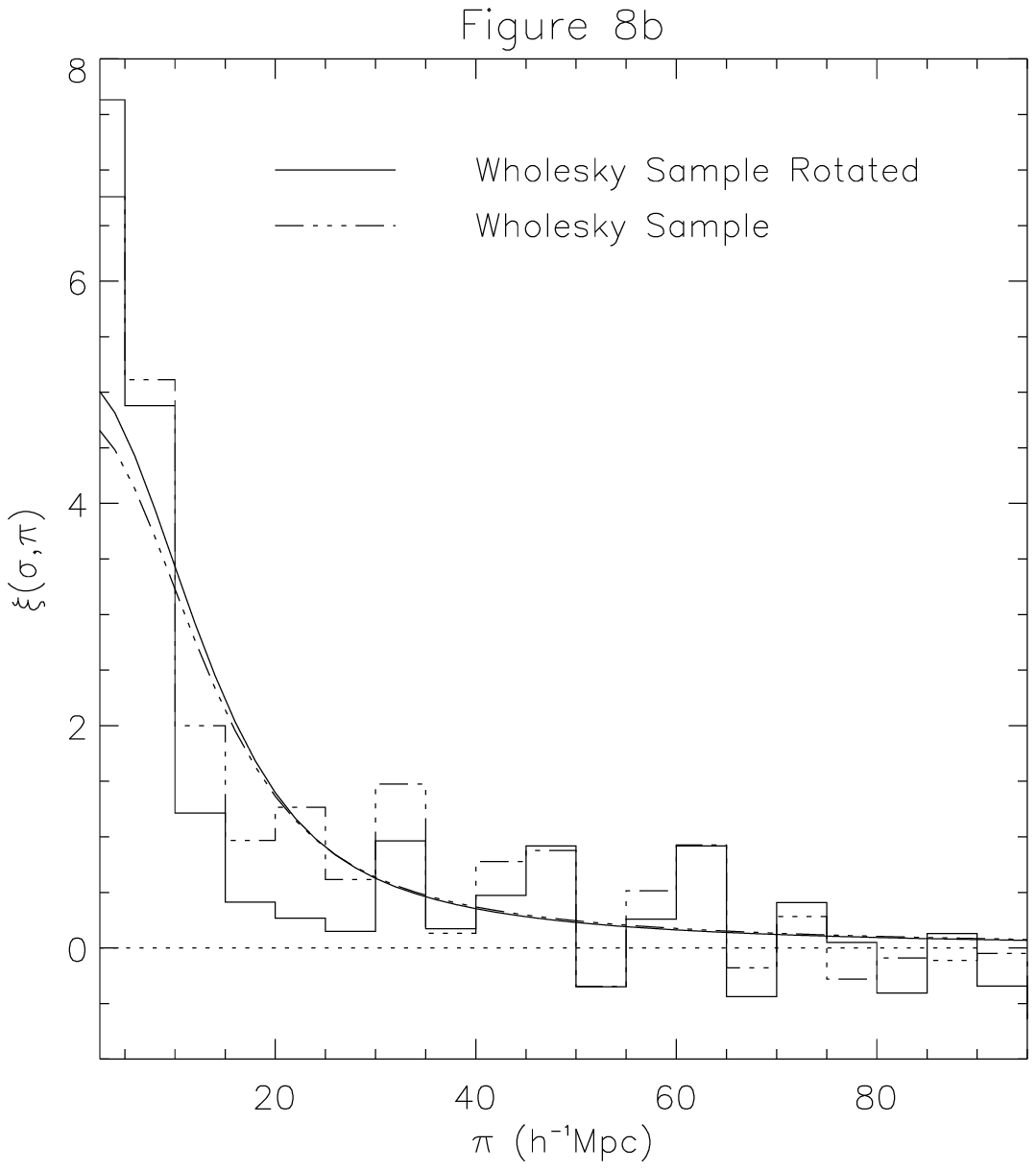}
\caption[]{
{\bf Fig 8a:} $\xi(\sigma,\pi)$ constrained to $0 \le \sigma \le 15 h^{-1}$Mpc 
for the PHG statistical sample (solid) and the whole-sky magnitude-limited
sample (dash-dot). The smooth line is the respective histogram convolved
with a Gaussian of width 700 km s$^{-1}$.  
{\bf Fig 8b:} Similar to Fig 8a. The solid line is for the magnitude-limited
sample after rotating the Ursa Majoris and Corona Borealis superclusters.
The dash-dot
line is for before the rotation.
}
\end{figure}


\begin{thebibliography}{}
\bibitem[(Abadi {\it et al.} 1998)]{aba98} Abadi, M., Lambas, D., and Muriel, H.
  1998, \apj, 507, 526
\bibitem[(Abell 1958)]{abe58} Abell, G. O.  1958, \apjsupp, ~ 3, 211
\bibitem[(Abell et al. 1989)]{aco89} Abell, G. O.,
  Corwin, H. G., Olowin, R. P. 1989, \apjsupp, ~70, 1 (ACO)
\bibitem[(Bahcall \& Cen, 1994)]{bac94} Bahcall, N.A. \& Cen, R. 1994,
  \apjlett, ~426, L15
\bibitem[(Bahcall \& Oh 1996)]{bao96} Bahcall, N.A., Oh, S.P. 1996,
  \apjlett, 462, L49
\bibitem[(Bahcall \& Soniera 1983)]{bns83} Bahcall, N.,
   Soniera, R.  1983, \apj, ~270, 20
\bibitem[(Bahcall {\it et al.} 1986)]{bac86} Bahcall, N., Soneira, R.,
and Burgett, W. 1986, \apj, 311, 15
\bibitem[(Bahcall, \& West 1992)]{baw92} Bahcall, N.A. \& West, M.J.
  1992, \apj, ~392, 419
\bibitem[(Bardeen, Steinhardt, \& Turner 1983)]{bst83} Bardeen, J.M.,
  Steinhardt, P.J., \& Turner, M.S. 1983,
  Phys. Rev. D, ~28, 679
\bibitem[(Bardeen {\it et al.} 1987)]{bar87} Bardeen, J.P., Bond, J.R.,
  \& Efstathiou, G. 1987, \apj, ~321, 28
\bibitem[(Batuski \& Burns 1985)]{bbb85} Batuski, D. J.,
  Burns, J. O.  1985, \apj, ~299, 5
\bibitem[(Batuski {\it et al.} 1989)]{bat89} Batuski, D.J, Bahcall, N.A,
  Burns, J.O., \& Olowin, R. 1989, \apj, 341, 599
\bibitem[(Batuski {\it et al.} 1999)]{bat99} Batuski, D.J., Miller, C.J.,
   Slinglend, K.A., Balkowski, C., Maurogordato, S., Cayatte, V., Felenbok, P.,
   \& Olowin, R. 1999, Under revision for \apj
\bibitem[(Croft \& Efstathiou 1994)]{cro95} Croft, R.A.C., \&
  Efstathiou, G. 1994, \mnras, 267, 390
\bibitem[(Dalton {\it et al.} 1994)]{dal94} Dalton, G.B., Croft, R.A.C.,
  Efstathiou, G., Sutherland, W.J., Maddox, S.J., and Davis, M. 1994, \mnras,
  271, 47
\bibitem[(Davis {\it et al.} 1985)]{dav85} Davis, M., Efstathiou, G., Frenk, C.S.,
   \& White, S.D.M. 1985, \apj, ~292, 371
\bibitem[(Ebeling et al. 1997)]{ebe97} Ebeling, H., Edge, A.C.,
  Fabian, A.C., Allen, S.W., Crawford, C.S., Boehringer, H. 1997,
  \apjlett, ~479, L101
\bibitem[(Efstathiou {\it et al.} 1992)]{eft92} Efstathious, G., Dalton, G.B.,
  Maddox, S.J., \& Sutherland, W. 1992, \mnras, ~257, 125
\bibitem[(Geller \& Huchra 1989)]{gel89} Geller, M., \& Huchra, J.P.
  1989, Science, ~246, 897
\bibitem[(Guth \& Pi 1982)]{gup82} Guth, A.H., \& Pi, S-Y. 1982,
  Phys. Rev. Lett., ~49, 1110
\bibitem[(Hamilton 1993)]{ham93} Hamilton, A.J.S. 1993, \apj, ~417, 19
\bibitem[(Huchra {\it et al.} 1990)]{huc90} Huchra, J.P., Henry, J.P.,
  Postman, M., \& Geller, M.J. 1990, \apj, 365, 66
\bibitem[(Jaaniste {\it et al.} 1998)]{jan86} Jaaniste, J., Tago, E., Einasto, M.,
  Einasto, J., Andernach, H., Mueller, V. 1998, A \& A, 336, 35
\bibitem[(Jing {\it et al.} 1992)]{jin92} Jing, Y.P., Plionis, M.,
    Valdarnini, R. 1992, \apj, ~389, 499
\bibitem[(Kaiser 1985)]{kai84} Kaiser, N. 1984, \apj, ~284, 9
\bibitem[(Katgert {\it et al.} 1996)]{kat96} Katgert,P., Mazure, A., Perea, J.,
  den Hartog, R., Moles, M., Le Fevre, O., Dubath, P., Focardi, P., Rhee, G.,
  Jones, B., Escalera, E., Biviano, A., Gerbal, D., Giuricin, G. 1996,
  A \& A, ~310, 8
\bibitem[(Kolb \& Turner 1990)]{kot90} Kolb, E.W. \& Turner, M.S. 1990, The Early
  Universe, Addison-Wesley Publishing Company
\bibitem[(Landy \& Szalay 1993)]{lan93} Landy, S.D. \& Szalay, A.S. 1993,
  \apj, ~ 412, 64
\bibitem[(Linde 1982)]{lin82} Linde, A. 1982, Phys. Lett. B, ~108, 389
\bibitem[(Ling {\it et al.} 1986)]{lin86} Ling, E.N., Frenk, C.S., Barrow, J.D.
  1986, \mnras, 223, 21
\bibitem[(Lucey 1983)]{luc83} Lucey, J.R. 1983, \mnras, ~203, 33
\bibitem[(Lumsden {\it et al.} 1992)]{lum92} Lumsden, S.L., Nichol, R.C., Collins, C.A.,
   \& Guzzo, L. 1992, \mnras, ~258, 1
\bibitem[(Maddox {\it et al.} 1990a)]{ma90a} Maddox, S.J., Sutherland, W.J.,
  Efstathiou, G., Loveday, J. 1990a, \mnras, ~243, 692
\bibitem[(Maddox {\it et al.} 1990b)]{ma90b} Maddox, S.J., Efstathiou, G.,
  Sutherland, W.J. 1990b, \mnras, ~246, 443
\bibitem[(Mazure {\it et al.} 1996)]{maz96} Mazure, A. {\it et al.} 1996, \mnras, ~310, 31
\bibitem[(Miller {\it et al.} 1999)]{mil99} Miller, C.J., Slinglend, K.A.,
  Batuski, D.J., \& Hill J.M. 1999, in preparation
\bibitem[(Narlikar 1983)]{nar83} Narlikar, J.V. 1983, Introduction to Cosmology,
  Cambridge University Press
\bibitem[(Peacock \& West 1992)]{pew92} Peacock, J.A. \& West, M.J. 1992,
  \mnras, 259, 494
\bibitem[(Nichol {\it et al.} 1992)]{nic92} Nichol, R.C., Collins, C.A.,
    Guzzo, L., \& Lumsden, S.L. 1992, \mnras, ~255, 21
\bibitem[(Nichol {\it et al.} 1994)]{nic94} Nichol, R.C., Briel, U.G., \&
  Henry, J.P. 1994, \mnras, ~267, 771
\bibitem[(Postman et al. 1986)]{pos86}  Postman, M.,
  Huchra, J. P., \& Geller, M. J.  1986, \aj, ~92, 1238 
\bibitem[(Postman et al. 1988)]{pos88}  Postman, M.,
   Geller, M. J., \& Huchra, J.P.  1988, \aj, ~95, 267 
\bibitem[(Postman et al. 1992)]{pos92}  Postman, M.,
  Huchra, J. P., \& Geller, M. J.  1992, \apj, ~384, 404
\bibitem[(Quintana \& Ramirez 1995)]{qur95} Quintana, H. \& Ramirez, A. 1995,
  \apjsupp, ~96, 343
\bibitem[(Ratcliffe {\it et al.} 1998)]{rat98} Ratcliffe, A., Shanks, T., 
    Parker, Q. \& Fong, R. 1998, \mnras, ~296, 173. 
\bibitem[(Sandage, 1975)]{san75} Sandage, A. 1975, Stars and Stellar Systems:
  Galaxies and the Universe, University of Chicago Press
\bibitem[(Schombert \& West 1990)]{shw90} Schombert, J.M. \& West, M.J.
  1990, \apj, ~363, 331
\bibitem[(Schuch 1981)]{sch81} Schuch, N. 1981, \mnras, ~196, 695S
\bibitem[(Schuch 1983)]{sch83} Schuch, N. 1983, \mnras, ~204, 1245S
\bibitem[(Shectman et al. 1996)]{she96} Shectman, S.A., et al. 1996, \apj,
  ~470, 172
\bibitem[(Slinglend {\it et al.} 1998)]{sli98} Slinglend, K.A., Batuski, D.J,
  Miller, C.M., Haase. S., Michaud, K., \& Hill, J.M. 1998, \apjsupp, ~115, 1
\bibitem[(Small {\it et al.} 1997a)]{sm97a} Small. T., Sargent, W., \& Hamilton,
   D. 1997, \apjsupp, ~111, 1
\bibitem[(Small {\it et al.} 1997b)]{sm97b} Small. T., Sargent, W., \& Hamilton,
   D. 1997, \apj, ~487, 512
\bibitem[(Small {\it et al.} 1998)]{sml98} Small. T., Ma, C., Sargent, W., \& Hamilton,
   D. 1997, \apj, ~492, 45 
\bibitem[(Struble \& Ftaclas, 1994)]{stf94} Struble, M.F \& Ftaclas, C.
  \aj, ~108, 1
\bibitem[(Struble \& Rood 1987)]{str87} Struble, M.F. \& Rood, H.J. 1987,
  \apjsupp, ~63, 543
\bibitem[(Struble \& Rood 1991)]{str91} Struble, M.F. \& Rood, H.J. 1991,
  \apj, ~374, 395
\bibitem[(Sutherland 1988)]{sut88} Sutherland, W. 1988, \mnras, ~234, 159
\bibitem[(Szalay \& Schramm 1985)]{sza85} Szalay, A. \& Schramm, D. 1985, Nature, ~
  ~314, 718
\bibitem[(Turok, 1989)]{tur89} Turok, N. 1989, Phys. Rev. Lett., 63, 2625
\bibitem[(Vilenkin, A. 1981)]{vil81} Vilenkin, A. 1981, Phys. Rev. Lett.,
  ~46, 1169
\bibitem[(West \& van den Bergh 1991)]{wev91} West, M.J., and van den Bergh, S.
  \apj, 373, 1
\bibitem[(WHite {\it et al.} 1987)]{whi87} White, S.D.M., Davis, M.,
    Efstathiou, G., \& Frenk, C.S. 1987, Nature, 330, 451
\bibitem[(Willmer {\it et al.} 1998)]{wil98} Willmer, C., Da Costa, L.N.,
  \& Pellegrini, P.S. 1998, \aj, ~115, 869
\bibitem[(Zabludoff {\it et al} 1993)]{zab93} Zabludoff, A., Geller, M., Huchra, J.P.,
   \& Vogeley, M. 1993, \aj, ~106, 1273
\bibitem[(Zeldovich 1980)]{zel80} Zeldovich, Ya.B., 1980, \mnras, ~192, 663
\bibitem[(Zucca {\it et al.} 1993)]{zuc93} Zucca, E., Zamorani, G., Scaramella, R.,
  \& Vettolani, G. 1993, \apj, ~407, 470
\end{thebibliography}
\end{document}